\documentclass[%
twocolumn,
superscriptaddress,
amsmath,amssymb,
aps,prx,
]{revtex4-2}
\usepackage{tabularx}
\usepackage{ltablex}
\usepackage{multirow}
\usepackage{amssymb,amsmath}
\usepackage{braket}
\usepackage[dvipsnames]{xcolor}
\usepackage{xcolor}
\usepackage{graphicx} 
\usepackage{dcolumn} 
\usepackage{bm} 
\usepackage{lipsum}

\usepackage{verbatim}
\usepackage{tikz}
\usepackage[compat=1.1.0]{tikz-feynman}
\tikzfeynmanset{warn luatex=false}
\usetikzlibrary{decorations.pathmorphing, decorations.markings, arrows.meta}

\usepackage[colorlinks=true,
            linkcolor=blue,
	        citecolor=blue,
	        urlcolor=blue]{hyperref}

\newcommand{\units}{Dipartimento di Fisica, Universit\`a di Trieste, I-34151 Trieste, Italy}
\newcommand{\sissa}{Scuola Internazionale Superiore di Studi Avanzati (SISSA), I-34136 Trieste, Italy}
\newcommand{\caltech}{Department of Applied Physics and Materials Science, California Institute of Technology, Pasadena, California 91125, USA}

\begin{document}

\title{First-principles real-space embedding theory of the superconducting proximity effect}
\author{Nicolas Baù}
\email{nicolas.bau@phd.units.it}
\affiliation{\units}
\author{Mitra Dowlatabadi}
\affiliation{\units}
\author{Tommaso Chiarotti}
\affiliation{\caltech}
\author{Massimo Capone}
\affiliation{\sissa}
\author{Antimo Marrazzo}
\email{amarrazz@sissa.it}
\affiliation{\sissa}

\date{\today} 
\begin{abstract}
    When a superconductor is placed in contact with a normal material, Cooper pairs penetrate the latter and induce superconductivity via the proximity effect. Despite its central role in quantum materials, superconducting devices and topological platforms, a predictive first-principles description of the proximity effect at realistic interfaces has remained computationally prohibitive so far. Here, we fill this gap by developing a Green's-function framework based on real-space dynamical embedding that enables first-principles simulations of superconducting proximity in mesoscopic systems. We show that the proximity effect admits a transparent diagrammatic formulation in terms of normal and anomalous embedding self-energies, which disentangle and quantify the distinct renormalization mechanisms generated by coupling to a superconducting bath. By combining this formalism with recursive schemes, we compute local spectral functions and proximity lengths extending over hundreds of nanometers into the bulk without resorting to thick interface slabs. We deploy the approach on tight-binding models (Qi-Hughes-Zhang and Fu-Kane-Mele), where we analyze mixed-parity superconductivity in topological insulators proximitized by $s$-wave superconductors, and on first-principles simulations of NbSe$_2$/CrBr$_3$ heterostructures based on density-functional theory and maximally-localized Wannier functions, the latter enabling direct comparison with scanning tunneling spectroscopy experiments. Our work provides a scalable and conceptually unified framework that bridges microscopic electronic structure and mesoscale proximity physics, enabling predictive atomistic simulations of superconducting interfaces.
\end{abstract}

\maketitle

\section{Introduction}
    When a normal (i.e., non-superconducting) material is placed in contact with a superconductor (SC), superconducting correlations can extend through the interface into the normal material. This phenomenon is known as the superconducting proximity effect (SPE) and it is often described in terms of Andreev reflections~\cite{meissner_1932, andreev_1964}. These are processes in which a Cooper pair is allowed to tunnel from the SC into the normal material\textemdash or equivalently, an electron is allowed to tunnel from the normal material into the SC forming a Cooper pair\textemdash by reflecting back a hole. The superconductivity induced by the SPE is substantially different from that of intrinsic SCs, as the anomalous density of Cooper pairs is finite even though locally there are no microscopic interactions pairing electrons.

    While the SPE has been extensively studied in many contexts\textemdash ranging from spintronics applications~\cite{linder_2015}, to the study of unconventional superconductivity~\cite{linder_rmp2019}, and the realization of Josephson junctions~\cite{golubov_rmp2004}\textemdash it has recently emerged as a central tool in the search for topological superconductivity~\cite{qi_rmp2011, alicea_2012}. Topological superconductors (TSC) exhibit gapless excitations that behave like Majorana fermions~\cite{majorana_1937} and are protected by a bulk topological invariant. At variance with topological insulators, these modes can either occur at the boundary (e.g., chiral Majorana fermions) or in superconducting vortices in the bulk. Particularly interesting are Majorana zero modes (MZM), localized zero-energy states that are promising candidates to realize anyons with non-Abelian statistics, the building block of fault-tolerant topological quantum computing~\cite{sarma_npj2015,yazdani_science2023}. MZMs are predicted to occur at the edge of one-dimensional (1D) TSCs~\cite{kitaev_2001}, in the vortices of 2D chiral TSCs~\cite{read_prb2000, sato_rpp2017}, or as corner modes in higher-order TSCs~\cite{khalaf_prb2018, zhu_prb2018, geier_prb2018, volpez_prl2019, zhu_prl2019, pahomi_prr2020, yang_jpcm2024}.

    Most proposals for realizing topological superconductivity and Majorana modes require either intrinsic or effective $p$-wave superconductivity~\cite{alicea_2012, bernevig_book2013, stanescu_book2024}. Nevertheless, despite the growing attention, intrinsic TSCs with $p$-wave pairing remain rare and elusive to identify. Phonon-mediated SCs are typically $s$-wave, while only strongly-correlated systems have shown some partial evidence of intrinsic $p$-wave pairing, with small superconducting gaps and very low critical temperatures~\cite{flototto_sciadv2018, mandal_cm2023}. Remarkably, $p$-wave topological superconductivity can emerge in heterostructures made of a conventional $s$-wave SC placed in contact with either a topological insulator or a semiconductor with Rashba spin texture, in presence of a Zeeman field; these are named artificial or designer TSCs~\cite{fu_prl2008, lutchyn_prl2010, qi_prb2010}. The Zeeman field can be either obtained through an external magnetic field or by proximity with a ferromagnet~\cite{alicea_2012}. In these systems, it is the underlying spin texture of the electronic band structure that enables effective triplet superconductivity induced by singlet correlations.The success of these proposals therefore hinges on an accurate description and control of the SPE, supported by reliable theoretical and computational frameworks.

    The SPE and artificial TSCs are often modeled by the mean-field Bogoliubov-de Gennes (BdG) approach, where the induced superconducting pairing $\Delta$ is introduced in the normal material as a fixed parameter~\cite{volkov_physica1995, fu_prl2008}, or it is computed self-consistently starting from some knowledge or parametrization of the microscopic electron-electron effective pairing interaction~\cite{wang_prb2024}.
    In general terms, but also to accurately describe real heterostructures, the induced superconductivity cannot be solely described by an additional pairing parameter leaving the normal-state Hamiltonian unchanged. This has been shown, for example, in Refs.~\cite{miller_pr1968, stanescu_prb2010, sau_prb2010, potter_prb2011}, where a surface self-energy derived from the coupling to a bulk BCS local Green's function is introduced in the Dyson equation of the renormalized system. Crucially, this self-energy contains both normal and anomalous components that are treated on equal footing, highlighting the role of normal-state renormalization that is often overlooked when superconductivity is modeled as a simple additional pairing term. More broadly, an unbiased treatment of frequency-dependent anomalous self-energies is also fundamental to describe superconductivity beyond the BCS-Migdal-Eliashberg treatment and reach the strong-coupling limit where strongly bound pairs are formed and condense, leading to a Bose-Einstein condensation regime~\cite{Micnas1990,ToschiPRB2005} which\textemdash for the electron-phonon case\textemdash is associated with the formation of bipolarons~\cite{Chakraverty1985,Capone2003,Zhang2023}.

    The methods discussed above are often based on simplified models that, while crucial for understanding general phenomena, lack the predictive power to discover and design materials\textemdash objectives that can instead be addressed by \textit{ab initio} approaches. An intermediate step towards realistic descriptions of specific materials and devices is offered by semi-phenomenological methods. These retain some simplifying assumptions\textemdash like effective Hamiltonians or parametrized superconducting gaps\textemdash while allowing system-dependent parameters to be adjusted in a more controlled way, for example by fitting to experimental data. For instance, in Ref.~\cite{hu_npj2025}, a predictive first-principles calculation for the normal electronic structure is combined with a phenomenological treatment of the proximity-induced superconductivity. This approach scales well with the size of the system, making it suitable for large-scale simulations, but still relies on a simplified description of the induced superconductivity, in which the decay of the induced superconducting gap follows a parametric equation~\cite{silvert_1975}. Another example, in the context of Josephson junctions, is the use of \emph{ab initio} tight-binding models based on Wannier functions combined with empirical pairing terms~\cite{chen_arxiv2025}.

    To overcome the limitations of such methods, pioneering efforts towards fully \textit{ab initio} simulation of the SPE have been mostly based on superconducting density functional theory (SCDFT)~\cite{luders_prb2005, marques_prb2005, linscheid_prb2015A, linscheid_prb2015B}. Here, the SPE is computed self-consistently with a generalization of the Kohn-Sham (KS) equations that include also the anomalous density. For instance, in Refs.~\cite{reho_prb2024, reho_prb2026}, the SPE is computed self-consistently by solving the spin-generalized KS-BdG equations with an initial guess for the superconducting order parameter in the SC. In a similar spirit, in Refs.~\cite{russmann_prb2022, russmann_arxiv2022, russmann_prr2023}, the predictive power of SCDFT is constrained by the need to assume a specific form for the superconducting pairing kernel and to choose superconducting coupling parameters~\cite{suvasini_prb1993}. A representative example is given in Ref.~\cite{russmann_prr2023}, where the induced pairing is obtained self-consistently by postulating an atom-dependent superconducting pairing kernel that is nonzero only in one of the materials forming the junction and is fitted to reproduce experimental data.

    Most current efforts to describe the SPE from first principles rely on SCDFT and address the problem by explicitly simulating a few layers of the interface. However, SCDFT faces intrinsic limitations that are related to the mismatch between the typical length/energy scales of the SPE and what can be simulated explicitly by first-principles electronic structure methods even on today's largest supercomputers. First and foremost, the proximity effects can manifest over mesoscopic length scales\textemdash often tens to hundreds of nanometres, i.e., up to hundreds of thousands of atoms. In addition, lattice mismatch between the two sides of the interface can require not only deep but also wide supercells. In other words, the proximity length is a mesoscopic scale far beyond what can be explicitly simulated \textit{ab initio}. Indeed, even when the interface is modeled with few tens or hundreds of atoms, SCDFT must be used semi-phenomenologically (i.e., with limited predictive power) due to the large computational cost, with a constant local pairing potential fitted to experiments rather than derived self-consistently~\cite{russmann_arxiv2022,russmann_prr2023, reho_prb2024,reho_prb2026}. In addition to the length scale, also the energy scale increases the computational cost: superconductivity occurs at the meV scale and typically needs very accurate grids in reciprocal space to properly integrate electron-phonon coupling and, more in general, other properties related to the pairing potential. Moreover, the framework of SCDFT requires all relevant materials, interfaces, and interactions to be described within a single SCDFT Hamiltonian, effectively limiting its flexibility and scalability when addressing complex heterostructures. Much of the reasoning outlined for SCDFT applies even more strongly to the principal alternative approach, i.e., many-body perturbation theory (MBPT) in the Migdal-Eliashberg approximation~\cite{margine_prb_2013,giustino_rmp_2017}—with the added drawback that no semi-phenomenological parametrization can be readily derived. While SCDFT and Migdal-Eliashberg are extremely powerful for studying the SPE, a fully \emph{ab initio} approach would likely need a more flexible framework that decouples the relevant energy and length scales\textemdash possibly incorporating SCDFT/MBPT calculations,  including dynamical mean-field theory (DMFT)~\cite{GeorgesRMP1996}, in a multi-stage workflow.

    Here, we present a comprehensive first-principles Green's function theory of the SPE based on real-space embedding potentials and Wannier functions. The effect of the SC on the normal material is encoded in two dynamical local self-energies\textemdash the embedding potentials\textemdash which renormalize both the normal and anomalous Green's functions of the proximitized material.

    \begin{figure*}
        \centering
        \includegraphics[width = \linewidth]{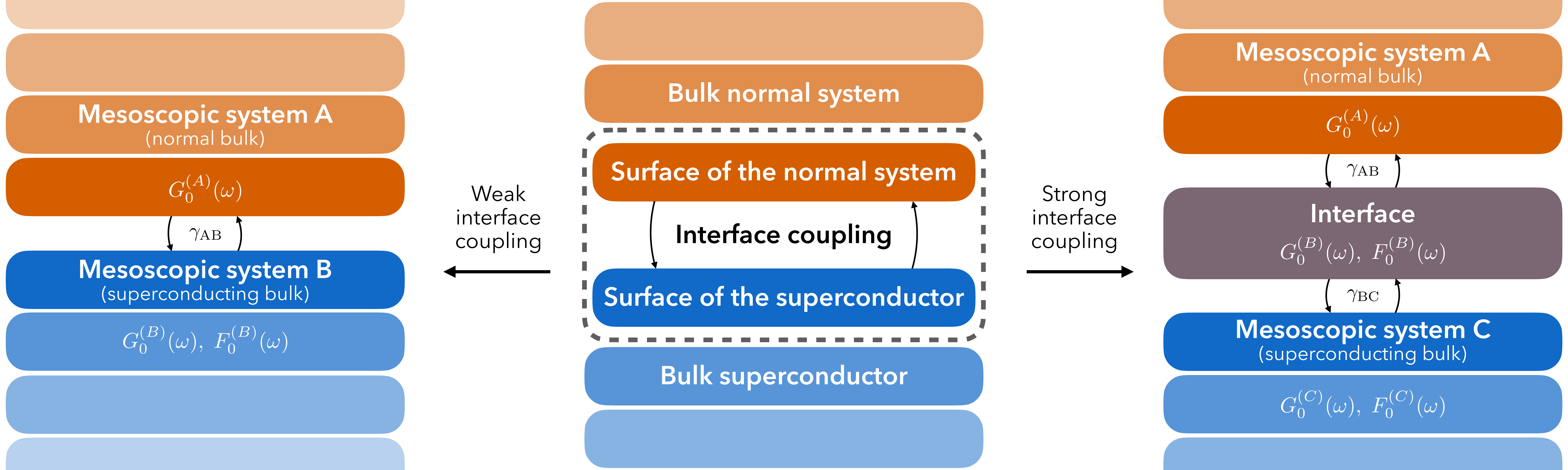}
        \caption{Schematic of a normal--superconductor (N/SC) heterostructure used in the embedding scheme. Depending on the strength of the interface coupling, two limits can be identified. If the interface coupling is weak (left panel), the two constituents of the heterostructure can be studied separately. In this case, the isolated N material (region \(A\)) is described by a Green's function \(G^{(A)}_{0}\), while the isolated SC (region \(B\)) is described by its normal and anomalous propagators, \(G^{(B)}_{0}\) and \(F^{(B)}_{0}\). Here, the SC (region \(B\)) plays the role of a bath and is formally integrated out, thereby effectively renormalizing the Green's functions of the N subsystem (region \(A\)). The two regions are coupled by a purely electronic single-particle term \(\gamma_{AB}\) (i.e., with no intrinsic superconducting pairing component in the coupling). If the interface coupling is strong (right panel), the interface (region \(B\)) explicitly contains portions of both materials and must therefore be solved as a standalone subsystem, and it is thus characterized by its normal and anomalous propagators \(G^{(B)}_{0}\) and \(F^{(B)}_{0}\). The bulk N system (region \(A\)) is described by the Green's function \(G^{(A)}_{0}\), while the bulk SC (region \(C\)) is described by the propagators \(G^{(C)}_{0}\) and \(F^{(C)}_{0}\). In the embedding picture the bulk N system (region \(A\)) and SC (region \(C\)) act as baths that are formally integrated out, thereby renormalizing the Green's functions of the explicit interface (region \(B\)). The couplings \(\gamma_{AB}\) and \(\gamma_{BC}\) that connect \(A\)\(\leftrightarrow\)\(B\) and \(B\)\(\leftrightarrow\)\(C\) are purely single-particle electronic hopping terms (i.e., they contain no intrinsic superconducting pairing components).}
        \label{fig:scheme}
    \end{figure*}

    The paper is organized as follows. In Sec.~\ref{sec:ab_initio_spe}, we develop the Green's function theory of the SPE with embedding potentials. We derive the Dyson equations for the SPE along with their diagrammatic representation. Then we extend the approach to mesoscopic systems through surface embedding and recursive schemes, and discuss the weak and strong limits of the interface coupling in the context of first-principles simulations. In Sec.~\ref{sec:numerical_results}, we validate and deploy our approach through numerical simulations within the BdG description of superconductivity. First, in Sec.~\ref{sec:qhz} we consider the Qi-Hughes-Zhang (QHZ) model~\cite{qi_prb2010} for chiral TSCs to compare our method to the standard BdG calculations with a fixed $s$-wave pairing. Then, in Sec.~\ref{sec:semiinfinite} we consider the SPE between mesoscopic (semi-infinite) systems: first an $s$-wave SC in contact with a single-band free-electron-like metal, and then with a three-dimensional (3D) topological insulator through the Fu-Kane-Mele (FKM) model~\cite{fu_prl2007}. We calculate the proximity lengths of the induced singlet and triplet anomalous densities, as well as local spectral densities. In Sec.~\ref{sec:scti}, we study the NbSe$_2$/CrBr$_3$ van der Waals heterostructure from first principles, including density functional theory, and compare with experiments. Finally, in Sec.~\ref{sec:conclusions}, we summarize our results and provide an outlook for future developments and applications.

\section{Embedding theory for superconducting interfaces}\label{sec:ab_initio_spe}
    Let us consider an interface composed of a normal system ($A$) and a SC ($B$), which are coupled by some single-particle electronic Hamiltonian ($\gamma_{AB}$), as represented schematically in Fig.~\ref{fig:scheme}.
    We call the Hamiltonian of the full ($A$+$B$) system $\mathcal H$ and its Green's function $\mathcal G(\omega)$, defined by $\left(\omega - \mathcal H\right)\mathcal G(\omega) = \mathbb I$. Following the notation of Ref.~\cite{marrazzo_rmp_2024}, we introduce projectors onto Wannier functions~\cite{marzari_prb_1997, marzari_rmp_2012,marrazzo_rmp_2024} $\ket{\mathbf{R}n}$ localized on either side of the interface as $\mathcal P_{\alpha}=\mathcal C\otimes \sum_{(\mathbf R, n)\in\alpha}|\mathbf Rn\rangle\langle\mathbf Rn|$, where $\mathcal C$ is the particle-hole operator. Then, we can partition the system into two subsystems $A$ and $B$ by applying the projectors to the total Hamiltonian, yielding the block structure:
    \begin{align}\begin{aligned}
        \label{eq:blocks}
        \mathcal H &_{\mathrm{BdG}} = \begin{pmatrix}
            \mathcal H^{(AA)}_{\mathrm{BdG}} & \mathcal H^{(AB)}_{\mathrm{BdG}} \\
            \mathcal H^{(AB)\dagger}_{\mathrm{BdG}} & \mathcal H^{(BB)}_{\mathrm{BdG}}
        \end{pmatrix}=\\[2mm] &=
        {\small
        \begin{pmatrix}
            h^{(A)} & 0 & h^{(AB)} & 0 \\
            0 & -h^{(A)*} & 0 & -h^{(AB)*} \\
            h^{(AB)\dagger} & 0 & h^{(B)} & \underline{\Delta}^{(B)} \\
            0 & -h^{(AB)T} & -\underline{\Delta}^{(B)*} & -h^{(B)*}
        \end{pmatrix}.
        }
    \end{aligned}\end{align}
    Here, $h$ and $\underline{\Delta}$ are respectively the normal and superconducting components of the projected BdG Hamiltonian $\mathcal H_{\mathrm{BdG}}^{(\alpha\beta)}$, defined as
    \begin{align}\label{eq:projection}
        \mathcal H_{\mathrm{BdG}}^{(\alpha\beta)}=\mathcal P_{\alpha}\mathcal H_{\mathrm{BdG}}\mathcal P_{\beta}.
    \end{align}

    In the following we only assume that the Hamiltonian coupling the two subsystems is static. For the sake of simplicity, since we target first-principles simulations, we write the following equations for static quasiparticle Hamiltonians, including KS DFT and MBPT (e.g., GW). We emphasize that the formalism that we derive in this section is more general and can naturally accommodate dynamical self-energies (both normal and anomalous) with an arbitrary frequency dependence for both subsystems.

    To keep the notation compact, in the following we will remove the double superscript in the projected Hamiltonians and rename the hopping term $AB$ as $\Gamma$ (and $AA$ as $A$, $BB$ as $B$). Similarly to what has been done for the Hamiltonian, the projection yields the total Green's function in a block form
    \begin{align}
        \mathcal G(\omega)=\begin{pmatrix}
            \mathbb G^{(A)}(\omega) & \mathbb G^{(\Gamma)}(\omega) \\
            \mathbb G^{(\Gamma)\dagger}(\omega) & \mathbb G^{(B)}(\omega)
        \end{pmatrix}
    \end{align}
    with the projected Green's functions $\mathbb G$ written in the Nambu formalism
    \begin{align}
        \mathbb G(\omega)=\begin{pmatrix}
            G(\omega) & F(\omega) \\
            F^{\dagger}(\omega) & -G^{T}(-\omega)
        \end{pmatrix}
    \end{align}
    where $G$ and $F$ are respectively the normal and anomalous Green's functions. It is important to note that owing to the real-space localization of the Wannier basis, the real-space partitioning can be equivalently done in reciprocal space, as explained in Appendix~\ref{app:wannier}. This allows us to write the secular problem as a set of coupled equations for the Green's function of the two subsystems.

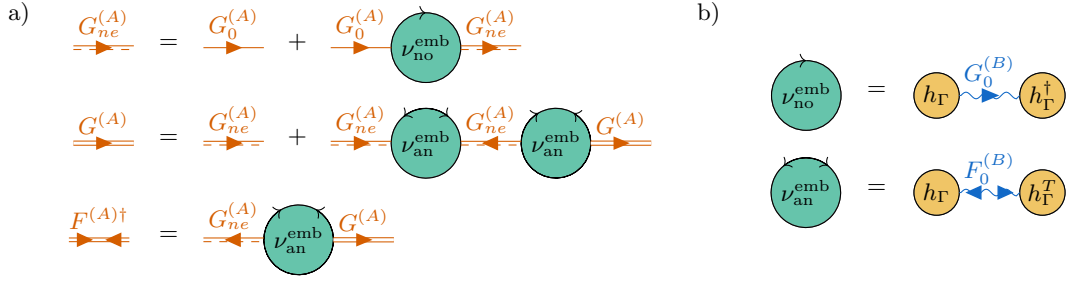
\begin{figure*}
        \centering
        \hfill \begin{minipage}{0.05\textwidth}
            \vspace{-3cm} a)
        \end{minipage}
        \begin{minipage}{0.25\textwidth}
            \centering
            \definecolor{potentials}{HTML}{009E73}
\definecolor{systa}{HTML}{D55E00}
\definecolor{systb}{HTML}{1269C7}
\definecolor{hgamma}{HTML}{E69F00}

\begin{equation*}\begin{aligned}
    \begin{tikzpicture}[baseline=(current bounding box.base)]\begin{feynman}
        \vertex (f1) at (-0.8,0.075em);
        \vertex (f2) at (0,0.075em);
        \vertex (f1m) at (-0.8,0);
        \vertex (f2m) at (0,0);

        \diagram* {
            (f1) -- [-, edge label = \(G^{(A)}_{ne}\), systa] (f2)
        };

        \draw[dashed, systa] (f1) ++(0,-0.15em) -- ++(0.8,0);
        
        \path[systa, postaction={decorate, decoration={markings, mark=at position 0.64 with {\arrow[scale=1.3]{Triangle[length = 1.65mm]}},}}] (f1m) -- (f2m);
    \end{feynman}\end{tikzpicture}
    \hspace{2mm}&=\hspace{2mm}
    \begin{tikzpicture}[baseline=(current bounding box.base)]\begin{feynman}
        \vertex (f1) at (0,0);
        \vertex (f2) at (0.8,0);

        \diagram* {
            (f1) -- [fermion, arrow size=0.15em, edge label=\(G^{(A)}_0\), systa] (f2)
        };
    \end{feynman}\end{tikzpicture}
    \hspace{2mm}+\hspace{2mm}
    \begin{tikzpicture}[baseline=(current bounding box.base)]\begin{feynman}
        \vertex (f1) at (-1.26,0);
        \vertex (f2) at (1.26,0.075em);
        \vertex (b2r) at (-0.46,0);
        \vertex (b2) at (0.46, 0.075em);
        \vertex (f1m) at (0.46,0);
        \vertex (f2m) at (1.46,0);

        \vertex (b) at (0,0);
        \draw[fill=potentials, opacity=0.6] (b) ++(0,0.46) arc[start angle=90, end angle=450, radius=0.46];
        \draw[<-] (b) ++(0,0.46) arc[start angle=90, end angle=450, radius=0.46];
        \vertex [circle, minimum size=0.92cm] (b) at (0,0) {\(\nu_{\mathrm{no}}^{\mathrm{emb}}\)};

        \diagram* [horizontal=f1 to f2, layered layout]{
            (f1) -- [fermion, arrow size=0.15em, edge label=\(G^{(A)}_0\), systa] (b2r),
            (b2) -- [plain, edge label = \(G^{(A)}_{ne}\), systa] (f2)
        };

        \draw[dashed, systa] (b) ++(0.46,-0.075em) -- ++(0.8,0);
        
        \path[systa, postaction={decorate, decoration={markings, mark=at position 0.5 with {\arrow[scale=1.3]{Triangle[length = 1.65mm]}},}}] (f1m) -- (f2m);
    \end{feynman}\end{tikzpicture}
    \\
    \begin{tikzpicture}[baseline=(current bounding box.base)]\begin{feynman}
        \vertex (f1) at (-0.8,0);
        \vertex (f2) at (0,0);

        \diagram* {
            (f1) -- [double, double distance=0.3ex,with arrow=0.5,arrow size=0.15em, edge label=\(G^{(A)}\), systa] (f2)
        };
    \end{feynman}\end{tikzpicture}
    \hspace{2mm}&=\hspace{2mm}
    \begin{tikzpicture}[baseline=(current bounding box.base)]\begin{feynman}
        \vertex (f1) at (-0.8,0.075em);
        \vertex (f2) at (0,0.075em);
        \vertex (f1m) at (-0.8, 0);
        \vertex (f2m) at (0,0);

        \diagram* {
            (f1) -- [plain, edge label = \(G^{(A)}_{ne}\), systa] (f2)
        };

        \draw[dashed, systa] (f1) ++(0,-0.15em) -- ++(0.8,0);
        
        \path[systa, postaction={decorate, decoration={markings,mark=at position 0.65 with {\arrow[scale=1.3]{Triangle[length = 1.65mm]}},}}] (f1m) -- (f2m);
    \end{feynman}\end{tikzpicture}
    \hspace{2mm}+\hspace{2mm}
    \begin{tikzpicture}[baseline=(current bounding box.base)]\begin{feynman}
        \vertex (f1) at (-1.26,0);
        \vertex (f1u) at (-1.26,0.075em);
        \vertex (f2) at (2.98,0);
        \vertex (f2r) at (1.26,0.075em);
        \vertex (b2r) at (-0.46,0.075em);
        \vertex (b20) at (-0.46,0);
        \vertex (b2) at (0.46, 0.075em);
        \vertex (f1m) at (0.46,0);
        \vertex (f2m) at (1.46,0);

        \vertex (b1) at (0,0);
        \draw[fill=potentials, opacity=0.6] (b1) ++(0,0.46) arc[start angle=90, end angle=450, radius=0.46];
        \draw[<->] (b1) ++(0.29568,0.35) arc[start angle=50, end angle=490, radius=0.46];
        \vertex [circle, minimum size=0.92cm] (b1) at (0,0) {\(\nu_{\mathrm{an}}^{\mathrm{emb}}\)};

        \vertex (b2c) at (1.26,0);
        \draw[fill=potentials, opacity=0.6] (b2c) ++(0.46,0.46) arc[start angle=90, end angle=450, radius=0.46];
        \draw[<->] (b2c) ++(0.23,0.39837) arc[start angle=120, end angle=780, radius=0.46];
        \vertex [circle, minimum size=0.92cm] (b2cr) at (1.72,0) {\(\nu_{\mathrm{an}}^{\mathrm{emb}}\)};
        \vertex (f2rr) at (2.18,0);

        \diagram* [horizontal=f1 to f2, layered layout]{
            (f1u) -- [plain, arrow size=0.15em, edge label=\(G^{(A)}_{ne}\), systa] (b2r),
            (b2) -- [plain, edge label = \(G^{(A)}_{ne}\), systa] (f2r),
            (f2rr) -- [double, double distance=0.3ex,with arrow=0.5,arrow size=0.15em, edge label=\(G^{(A)}\), systa] (f2)
        };

        \draw[dashed, systa] (b) ++(0.46,-0.075em) -- ++(0.8,0);
        \draw[dashed, systa] (f1) ++(0,-0.075em) -- ++(0.8,0);

        \path[systa, postaction={decorate, decoration={markings, mark=at position 0.72 with {\arrow[scale=1.3]{Triangle[length = 1.65mm]}},}}] (f2m) -- (f1m);
        \path[systa, postaction={decorate, decoration={markings, mark=at position 0.65 with {\arrow[scale=1.3]{Triangle[length = 1.65mm]}},}}] (f1) -- (b20);
    \end{feynman}\end{tikzpicture}
    \\
    \begin{tikzpicture}[baseline=(current bounding box.base)]\begin{feynman}
        \vertex (f1) at (-0.8,0.075em);
        \vertex (f2) at (0,0.075em);
        \vertex (f1m) at (-0.8,0);
        \vertex (f2m) at (0,0);

        \diagram* {
            (f1) -- [plain, arrow size=0.15em, edge label=\(F^{(A)\dagger}\), systa] (f2)
        };

        \draw[plain, systa] (f1) ++(0,-0.15em) -- ++(0.8,0);
        
        \path[systa, postaction={decorate, decoration={markings,mark=at position 0.4 with {\arrow[scale=1.3]{Triangle[length = 1.65mm]}},}}] (f1m) -- (f2m);
        \path[systa, postaction={decorate, decoration={markings,mark=at position 0.6 with {\arrowreversed[scale=1.3]{Triangle[length = 1.65mm]}},}}] (f1m) -- (f2m);
    \end{feynman}\end{tikzpicture}
    \hspace{2mm}&=\hspace{2mm}
    \begin{tikzpicture}[baseline=(current bounding box.base)]\begin{feynman}
        \vertex (f1) at (-1.26,0);
        \vertex (f1u) at (-1.26,0.075em);
        \vertex (f2) at (2.98,0);
        \vertex (f2r) at (1.26,0.075em);
        \vertex (b2r) at (-0.46,0.075em);
        \vertex (b20) at (-0.46,0);
        \vertex (b2) at (0.46, 0.075em);
        \vertex (f1m) at (0.46,0);
        \vertex (f2m) at (1.46,0);

        \vertex (b2c) at (1.26,0);
        \draw[fill=potentials, opacity=0.6] (b2c) ++(0.46,0.46) arc[start angle=90, end angle=450, radius=0.46];
        \draw[<->] (b2c) ++(0.23,0.39837) arc[start angle=120, end angle=780, radius=0.46];
        \vertex [circle, minimum size=0.92cm] (b2cr) at (1.72,0) {\(\nu_{\mathrm{an}}^{\mathrm{emb}}\)};
        \vertex (f2rr) at (2.18,0);

        \diagram* [horizontal=f1 to f2, layered layout]{
            (b2) -- [plain, edge label = \(G^{(A)}_{ne}\), systa] (f2r),
            (f2rr) -- [double, double distance=0.3ex,with arrow=0.5,arrow size=0.15em, edge label=\(G^{(A)}\), systa] (f2)
        };

        \draw[dashed, systa] (b) ++(0.46,-0.075em) -- ++(0.8,0);

        \path[systa, postaction={decorate, decoration={markings, mark=at position 0.72 with {\arrow[scale=1.3]{Triangle[length = 1.65mm]}},}}] (f2m) -- (f1m);
    \end{feynman}\end{tikzpicture}
\end{aligned}\end{equation*}
        \end{minipage}
        \hfill\hfill
        \quad \begin{minipage}{0.05\textwidth}
            \vspace{-3cm} b)
        \end{minipage}
        \begin{minipage}{0.25\textwidth}
            \centering
            \definecolor{potentials}{HTML}{009E73}
\definecolor{systa}{HTML}{D55E00}
\definecolor{systb}{HTML}{1269C7}
\definecolor{hgamma}{HTML}{E69F00}
    
\begin{equation*}\begin{aligned}
    \begin{tikzpicture}[baseline=(current bounding box.base)]\begin{feynman}
        \vertex (b) at (0,0);

        \draw[fill=potentials, opacity=0.6] (b) ++(0,0.46) arc[start angle=90, end angle=450, radius=0.46];
        \draw[<-] (b) ++(0,0.46) arc[start angle=90, end angle=450, radius=0.46];
        \vertex [circle, minimum size=0.96cm] (b) at (0,0) {\(\nu_{\mathrm{no}}^{\mathrm{emb}}\)};
    \end{feynman}\end{tikzpicture}
    \hspace{2mm}&=\hspace{2mm}
    \begin{tikzpicture}[baseline=(current bounding box.base)]\begin{feynman}
        \vertex (f1) at (-0.4,0);
        \vertex (f2) at (0.4,0);
        
        \vertex (f0) at (-0.7,0);
        \draw[fill=hgamma, opacity=0.6] (f0) ++(0,0.3) arc[start angle=90, end angle=450, radius=0.3];
        \draw[-] (f0) ++(0,0.3) arc[start angle=90, end angle=450, radius=0.3];
        \vertex [circle, minimum size=0.6cm] (f0) at (-0.7, 0) {\(h_{\Gamma}\)};
        
        \vertex (f3) at (0.7,0);
        \draw[fill=hgamma, opacity=0.6] (f3) ++(0,0.3) arc[start angle=90, end angle=450, radius=0.3];
        \draw[-] (f3) ++(0,0.3) arc[start angle=90, end angle=450, radius=0.3];
        \vertex [circle, minimum size=0.6cm] (f3) at (0.7, 0) {\(h_{\Gamma}^{\dagger}\)};

        \diagram* {
            (f1) -- [charged boson, edge label = \(G^{(B)}_0\), systb] (f2),
        };
    \end{feynman}\end{tikzpicture}
    \\
    \begin{tikzpicture}[baseline=(current bounding box.base)]\begin{feynman}
        \vertex (b) at (0,0);

        \draw[fill=potentials, opacity=0.6] (b) ++(0,0.46) arc[start angle=90, end angle=450, radius=0.46];
        \draw[<->] (b) ++(0.29568,0.35) arc[start angle=50, end angle=490, radius=0.46];
        \vertex [circle, minimum size=0.96cm] (b) at (0,0) {\(\nu_{\mathrm{an}}^{\mathrm{emb}}\)};
    \end{feynman}\end{tikzpicture}
    \hspace{2mm}&=\hspace{2mm}
    \begin{tikzpicture}[baseline=(current bounding box.base)]\begin{feynman}
        \vertex (f1up) at (-0.4, 0.0);
        \vertex (f2up) at (0.4, 0.0);
        \vertex (f1) at (-0.4,0);
        \vertex (f2) at (0.4,0);
        
        \vertex (f0) at (-0.7,0);
        \draw[fill=hgamma, opacity=0.6] (f0) ++(0,0.3) arc[start angle=90, end angle=450, radius=0.3];
        \draw[-] (f0) ++(0,0.3) arc[start angle=90, end angle=450, radius=0.3];
        \vertex [circle, minimum size=0.6cm] (f0) at (-0.7, 0) {\(h_{\Gamma}\)};
        
        \vertex (f3) at (0.7,0);
        \draw[fill=hgamma, opacity=0.6] (f3) ++(0,0.3) arc[start angle=90, end angle=450, radius=0.3];
        \draw[-] (f3) ++(0,0.3) arc[start angle=90, end angle=450, radius=0.3];
        \vertex [circle, minimum size=0.6cm] (f3) at (0.7, 0) {\(h_{\Gamma}^{T}\)};

        \diagram* {
            (f1) -- [boson, systb, edge label = \(F^{(B)}_0\)] (f2)
        };

        \path[systb, postaction={decorate, decoration={markings,mark=at position 0.1 with {\arrowreversed[scale=1.3]{Triangle[length = 1.65mm]}},mark=at position 0.9 with {\arrow[scale=1.3]{Triangle[length = 1.65mm]}}}}] (f1) -- (f2);
    \end{feynman}\end{tikzpicture}
\end{aligned}\end{equation*}
        \end{minipage}
        \hfill\hfill
        \caption{
        Left panel (a): Diagrammatic representation of the embedding Dyson equations, Eqs.~\ref{eq:gel_embedding}--\ref{eq:induced_anomalous}, that describe how the normal Green's function \(G^{(A)}(\omega)\) and the induced anomalous Green's function \(F^{(A)\dagger}(\omega)\) of the normal subsystem \(A\) are renormalized by proximity to the superconductor \(B\).
        First, the normal propagator of \(A\) is renormalized by the normal part of \(B\), yielding the normally-embedded propagator \(G^{(A)}_{ne}(\omega)\) through the normal embedding potential \(\nu^{\mathrm{emb}}_{\mathrm{no}}(\omega)\); \(\nu^{\mathrm{emb}}_{\mathrm{no}}\) scales as \(h_{\Gamma}^{2}\) (where \(h_{\Gamma}\) is the single-particle interface coupling).
        Second, the fully renormalized \(G^{(A)}(\omega)\) is obtained from a Dyson equation for \(G^{(A)}_{ne}(\omega)\) with a self-energy that contains products of the anomalous embedding potential \(\nu^{\mathrm{emb}}_{\mathrm{an}}(\omega)\); because \(\nu^{\mathrm{emb}}_{\mathrm{an}}\) itself scales with \(h_{\Gamma}^{2}\), these contributions scale as \(h_{\Gamma}^{4}\).
        The same anomalous embedding terms are responsible for a non-vanishing induced anomalous propagator \(F^{(A)\dagger}(\omega)\) in the originally normal system.
        Right panel (b): Definitions of the normal and anomalous embedding potentials, \(\nu^{\mathrm{emb}}_{\mathrm{no}}(\omega)\) and \(\nu^{\mathrm{emb}}_{\mathrm{an}}(\omega)\), expressed in terms of the interface coupling \(h_{\Gamma}\) and the closed normal \(G^{(B)}_{0}(\omega)\) and anomalous  \(F^{(B)}_{0}(\omega)\) Green's functions of the isolated superconductor.}
        \label{fig:embedding_eqs}
    \end{figure*}

    Here, we focus on the renormalization of the Green's function of the normal system $A$ due to the proximity with the SC $B$, for which we can solve $(\omega-\mathcal H_{\mathrm{BdG}})\mathcal G(\omega)=\mathbb I$ to obtain
    \begin{widetext}
    \begin{align}\begin{cases}\label{eq:system_tosolve}
        [\omega - h^{(A)} - h^{(\Gamma)} G^{(B)}_0(\omega) h^{(\Gamma)\dagger}] G^{(A)}(\omega)  +  h^{(\Gamma)} F^{(B)}_0( \omega) h^{(\Gamma)T} F^{(A)\dagger}( \omega) = \mathbb{I} \\[3mm]
        h^{(\Gamma)*} F^{(B)\dagger}_0(\omega) h^{(\Gamma)\dagger} G^{(A)}(\omega) + [\omega + h^{(A)*} \ h^{(\Gamma)*} G^{(B)T}_0(-\omega) h^{(\Gamma)T}] F^{(A)\dagger}(\omega) = \mathbb{O}
    \end{cases}\end{align}
    \end{widetext}
    where $G^{(B)}_0(\omega)$ and $F^{(B)}_0(\omega)$ are the closed normal and anomalous Green's function of the isolated SC. Notably, the solution for $G^{(A)}(\omega)$ can be better understood by rewriting it as a two-step process. First, $G^{(A)}_0(\omega)=(\omega-h^{(A)})^{-1}$ is renormalized by the presence of the normal component of the SC:
    \begin{align}\label{eq:gel_embedding}
        G_{ne}^{(A)}(\omega)=G^{(A)}_0(\omega)+G^{(A)}_0(\omega)\nu^{\mathrm{emb}}_{\mathrm{no}}(\omega)G^{(A)}_{ne}(\omega)
    \end{align}
    where $\nu^{\mathrm{emb}}_{\mathrm{no}}(\omega)$ is the local embedding potential for the normal Green's function that is defined by
    \begin{align}\label{eq:embedding_el}
        \nu^{\mathrm{emb}}_{\mathrm{no}}(\omega) = h^{(\Gamma)}G^{(B)}_0(\omega)h^{(\Gamma)\dagger}.
    \end{align}
    Subsequently, the partially-renormalized (or ``normally-embedded'') Green's function $G^{(A)}_{ne}(\omega)$ undergoes a second renormalization due to the anomalous component of the SC Green's function $F^{(B)}_0(\omega)$:
    \begin{align}\begin{aligned}\label{eq:final_ga_embedding}
        &G^{(A)}(\omega)=G_{ne}^{(A)}(\omega)+\\ &\hspace{2mm} +G_{ne}^{(A)}(\omega)\nu^{\mathrm{emb}}_{\mathrm{an}}(\omega)G_{ne}^{(A)T}(-\omega)\nu_{\mathrm{an}}^{\mathrm{emb},\dagger}(\omega)G^{(A)}(\omega)
    \end{aligned}\end{align}
    where the anomalous embedding potential $\nu_{\mathrm{an}}^{\mathrm{emb}}(\omega)$ is defined as
    \begin{align}\label{eq:embedding_anom}
        \nu^{\mathrm{emb}}_{\mathrm{an}}(\omega) = h^{(\Gamma)}F^{(B)}_0(\omega)h^{(\Gamma)T}.
    \end{align}
    Finally, we can see that the normal system $A$ acquires also a non-vanishing anomalous Green's function given by
    \begin{align}\label{eq:induced_anomalous}
        F^{(A)\dagger}(\omega) = -G^{(A)T}_{ne}(-\omega)\nu^{\mathrm{emb},\dagger}_{\mathrm{an}}(\omega)G^{(A)}(\omega).
    \end{align}
    We diagrammatically represent Eqs.~\ref{eq:gel_embedding}--\ref{eq:induced_anomalous} in Fig.~\ref{fig:embedding_eqs}, making explicit the two distinct contributions to the SPE. In particular, we can see that $F^{(B)}$ does not renormalize the Green's function of system $A$ directly, but only through the normally-embedded Green's function $G^{(A)}_{ne}(\omega)$. This is a crucial, yet often overlooked, point: in general, the renormalization of the normal system\textemdash and the concomitant emergence of an induced anomalous component\textemdash cannot be faithfully captured by simply adding a constant pairing term to the Hamiltonian, as is often done in the literature. Our observation aligns with earlier studies based on simplified models~\cite{alicea_2012, miller_pr1968, stanescu_prb2010, sau_prb2010, potter_prb2011}, which already emphasized the importance of accounting for the normal component of the Hamiltonian, albeit within a different formalism.

    As reported in Appendix~\ref{app:triplet}, from the diagrams shown in Fig.~\ref{fig:embedding_eqs} we see that we can induce a non-vanishing triplet component in the normal system, for example, via spin-orbit interactions, producing off-diagonal terms in the Green's function, or via magnetic fields~\cite{bergeret_rmp2005}. Then, in general, a mixed-parity state comprising both singlet and triplet components with no net spin magnetization will appear in proximitized materials if spin-orbit coupling (SOC) is present~\cite{tkachov_quantummaterials}. It is worth mentioning that a non-vanishing triplet component can also arise in altermagnets despite the absence of SOC and external magnetic fields~\cite{ouassou_prl2023,li_prl2026}. It should also be noted that, while in disordered systems the $p$-wave component can be suppressed relative to the $s$-wave one, in clean systems the two can be comparable~\cite{tkachov_prb2013}.

    As already mentioned, it is worth emphasizing that the present framework relies essentially on the single approximation that the coupling between the two subsystems $A$ and $B$ is static and remains unrenormalized by the interactions. Furthermore, superconducting pairing is taken to occur only within each subsystem. In the present discussion, our focus is on the direct proximity effect, hence in Eq.~\ref{eq:system_tosolve} we considered the case where subsystem $A$ is not superconducting in the absence of system $B$. However, in Appendix~\ref{app:strongcoupling} we report the general equations when both systems can be intrinsically superconducting, providing a way to study also the inverse proximity effect, where a SC is renormalized by the presence of a normal material. We are going to use this more general set of equations in Sec.~\ref{sec:ab_initio_spe}.

    Due to their general character, these equations are not limited to a BdG or Migdal-Eliashberg framework, but they can be used also in combination with unbiased many-body approaches that allow computing the normal and anomalous self-energies for isolated systems. A natural candidate is DMFT~\cite{GeorgesRMP1996}, which enables an accurate treatment of the electron-electron correlations in both subsystems as well as describing superconductivity beyond the weak-coupling paradigm for both electronic and phononic pairings. DMFT and related methods have been successfully used to study superconductivity in models and in correlated materials including, e.g., alkali-metal doped fullerides~\cite{Nomura2015,Capone2009,Witt2024}, nickelates~\cite{Kitatani2020} and multilayer cuprates~\cite{BacqLabrueil2025}.
    A systematic study of all these applications defines a very broad research program that lies beyond the scope of the present work and is deferred to future investigation.

    \subsection{Surface embedding for mesoscopic systems\label{sec:LS}}
        In real interfaces, the SPE can extend over mesoscopic length scales\textemdash up to hundreds of nanometers~\cite{uday_nature2024}\textemdash while the induced superconducting gaps are typically small. As a consequence, explicit simulations of the interface would require simultaneously dense reciprocal-space grids and very thick slabs to reach convergence. To overcome these issues, we solve the embedding equations for surface Green's functions that describe the boundary of truly semi-infinite systems and are obtained by recursive embedding methods in the spirit of the López-Sancho iterative scheme~\cite{lopezsancho_jpf1984, lopezsancho_jpf1985}. López-Sancho schemes based on Wannier functions have been used for many applications, such as probing the existence of topological surface states or simulating electronic transport in lead-conductor-lead geometries~\cite{nardelli_prb_1999,calzolari_prb2004,lee_prl_2005,marrazzo_rmp_2024}. The Hamiltonian is partitioned into ``principal layers'' that are coupled only through nearest-layer hoppings. Then, an iterative algorithm is used to solve the secular equation $\left( \omega-\mathcal H \right)G(\omega)=\mathbb I$ through embedding self-energies (a.k.a. transfer matrices) and gives access to the surface Green's function from the sole knowledge of the bulk. This is particularly accurate in the limit where surface reconstruction and other self-consistent effects can be neglected.

        The adoption of the recursive scheme enables the study of the SPE in mesoscopic systems by solving the embedding equations of Fig.~\ref{fig:embedding_eqs} in terms of surface rather than bulk Green's functions, with the former computed via recursive techniques. In this way, a surface between two semi-infinite systems can be simulated explicitly without the need of calculating very thick slabs, which would be computationally prohibitive to converge properly. This becomes critical especially in systems hosting Majorana boundary modes: due to the small bulk SC gaps, these can hybridize across the slab if not sufficiently thick, thus opening spurious energy gaps at the boundary and compromising their topological properties. The issue appears twice in superconducting heterostructures with topological insulators, where the Dirac cones can exhibit spurious gaps due to finite thickness even before considering superconductivity.

    \subsection{Weak interface coupling limit}
        We emphasize that up to this point the equations are exact, in the sense that no approximations have been introduced beyond the assumption of a static coupling Hamiltonian for the materials forming the interface. In principle, Eq.~\ref{eq:projection} would require the knowledge of the Hamiltonian of the entire system $\mathcal{H}$, that is later to be projected on the manifolds of interest. For simple tight-binding or continuum models\textemdash and more in general any effective theory that does not depend on a global variable\textemdash this is what is actually done as we have easy access to $\mathcal{H}$. More in general, one wants precisely to avoid dealing explicitly with $\mathcal{H}$, which is typically a functional of a variable defined over the entire system. Key examples are DFT, where $\mathcal{H}$ is a functional of the total charge density $\rho$ or MBPT that requires the knowledge of the total Green's function. Nevertheless, even when a global variable is involved, the equations are still exact, provided they are solved self-consistently. In this case, the renormalization of the Green's function defines a new global variable (e.g. in DFT, a new $\rho$) that modifies the interface Hamiltonian and would normally require an iterative self-consistent treatment. For simplicity, here we identify two limiting cases in which the self-consistent cycle can be avoided, allowing the embedding equations to be solved in a single step, while leaving the fully self-consistent generalization for future work.

        If the interface coupling $h^{(\Gamma)}$ is weak, as is the case in van der Waals heterostructures, it is natural to approximate the projected interface Hamiltonians by those of the corresponding closed (isolated) systems. As already mentioned, while the use of closed-system quantities is exact within a tight-binding framework, it is, in general, an approximation. Focusing on the practically most relevant setting of DFT, where the Hamiltonian is a functional of the ground-state charge density, $\mathcal{H}[\rho]$, the weak-coupling approximation corresponds to
        \begin{align}\label{eq:weak_limit_def}
            \mathcal P_{\alpha}\mathcal H_{\mathrm{BdG}}[\rho]\mathcal P_{\alpha}\approx\mathcal H_{\mathrm{BdG}}[\rho_{\alpha}^{\mathrm{0}}],
        \end{align}
        where $\rho_{\alpha}^{\mathrm{0}}$ is the ground-state density of the closed (isolated) system $\alpha$. Here, the Green's-function calculations for the two subsystems are fully decoupled, allowing one to deploy different levels of theory on the two sides of the interface. A full electronic-structure calculation of the interface is still required, but only to extract the interfacial hopping amplitudes, thereby eliminating the need to explicitly simulate superconductivity across the interface.

    \subsection{Strong interface coupling limit}
        If the interface coupling is particularly strong, effects like band bending and charge transfer across the interface might not be sufficiently well described by the previous approximation, and Eq.~\ref{eq:weak_limit_def} could not hold anymore: the interface might need to be simulated explicitly and self-consistently converged.
        The simplest approach, sketched in Fig.~\ref{fig:scheme}, is to study a supercell composed of a thick finite slab of both materials, where the Hamiltonian of each subsystem can be computed via projection as in Eq.~\ref{eq:projection}. Still, we exploit recursive schemes to treat mesoscopic leads: if the slab is sufficiently thick that, far from the interface, the material is well approximated by its bulk, the semi-infinite geometry can be mimicked by embedding the interface on both sides. The required thickness is then substantially smaller than in the absence of embedding, where spurious surface states would appear on the two sides of the finite slab. This strong-coupling limit is in the spirit of Ref.~\cite{petocchi_prb2016} and is closer in philosophy to electronic-transport calculations in lead--conductor--lead geometries~\cite{nardelli_prb_1999,calzolari_prb2004,lee_prl_2005,marrazzo_rmp_2024}; it has also been employed to study Josephson currents in Josephson junctions~\cite{bansil_prb2023}. In this setting, we solve the embedding equations for three subsystems: two superconducting (the parent SC and the interface) and one normal material. The corresponding embedding solution is reported in Appendix~\ref{app:strongcoupling}. As mentioned previously, it should also be noted that Fig.~\ref{fig:scheme} admits a broader interpretation and can be adapted to study several related scenarios. These include, for instance, the inverse proximity effect, SC-SC junctions, and arbitrary multilayer heterostructures. The corresponding embedding solutions can be obtained by taking the appropriate limit starting from the equations developed in Appendix~\ref{app:strongcoupling}.

\section{Numerical results}\label{sec:numerical_results}
    Now, we validate and deploy the embedding theory developed in Sec.~\ref{sec:ab_initio_spe} via numerical simulations on four representative cases, ranging from minimal two-dimensional (2D) tight-binding models to first-principles simulations of real heterostructures. We begin with the QHZ model~\cite{qi_prb2010} for 2D quantum anomalous Hall insulators in proximity of $s$-wave SCs, where we benchmark the embedding solution against the standard BdG framework and compare with exact diagonalization methods. Then we consider mesoscopic systems and examine the SPE induced first in a 3D single-band free-electron-like metal and then in 3D topological insulators through the FKM model~\cite{fu_prl2007}. Here, we calculate the proximity lengths of singlet and triplet anomalous densities. Finally, we deploy our embedding theory in first-principles simulations of NbSe$_2$/CrBr$_3$ van der Waals heterostructures to investigate both the direct and inverse proximity effects, computing renormalized spectral functions as well as the layer-resolved density of states and superconducting gaps. There, we also comment on how our approach can aid the interpretation of experimental measurements.

    \subsection{2D quantum anomalous Hall insulators in proximity of \texorpdfstring{$s$}{}-wave superconductors}\label{sec:qhz}
\begin{figure*}
            \centering
            \includegraphics[width=\textwidth]{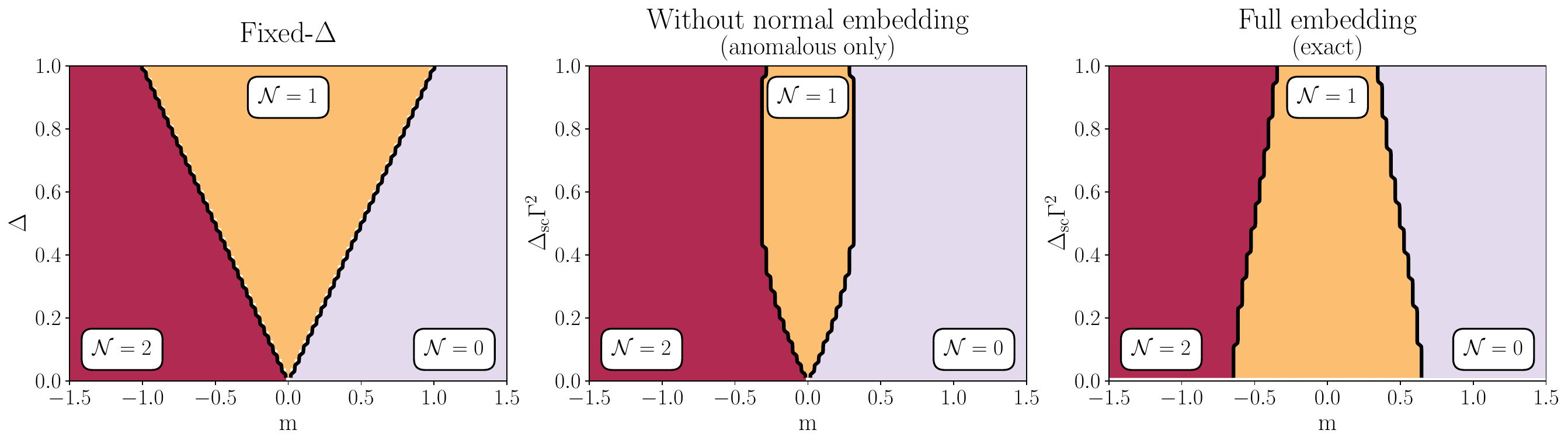}
            \caption{Topological phase diagrams of a quantum anomalous Hall insulator (QAHI) in proximity of an $s$-wave superconductor (SC), as a function of the QAHI mass parameter and of the induced pairing. In the fixed-$\Delta$ approach (left panel), a Bogoliubov-de Gennes Hamiltonian is constructed by considering the QHZ model and adding a constant pairing $\Delta$~\cite{qi_prb2010}. In the right-hand panel, we report the exact solution of the embedding equations for a QAHI/SC heterostructure, where $\Delta_{\mathrm{sc}}$ is the bulk paring of the SC and $\Gamma=0.8$ is the interface coupling. The center panel reports the solution of the embedding equations where the normal embedding potential is neglected (``anomalous only''). While three phases with distinct Chern numbers appear in all cases, the phase diagram can be rather different. Indeed, the fixed-$\Delta$ approach does not describe the renormalization of the normal Hamiltonian, as can be appreciated by the similarity with the anomalous-only embedding solution at small values of the induced pairings.}
            \label{fig:qhz_comparison}
        \end{figure*}

        The QHZ model~\cite{qi_prb2010} describes a 2D chiral TSC obtained by placing a quantum anomalous Hall insulator (QAHI) in proximity of an $s$-wave SC. The Hamiltonian of the QAHI is given by
        \begin{equation}\label{eq:qhz_hamilton}
            h_{\mathrm{QAHI}}(\mathbf p) = \begin{pmatrix}
                \epsilon_{\mathbf p} & A_{\mathbf p} \\
                A_{\mathbf p}^* & -\epsilon_{\mathbf p}
            \end{pmatrix},
        \end{equation}
        where $\epsilon_{\mathbf p}=m+Bp^2$ and $A_{\mathbf p}=A(p_x-ip_y)$, in the basis $\psi_{\mathbf p}=(c_{\mathbf p\uparrow},c_{\mathbf p\downarrow})$. The energy spectrum is gapped as long as $m\neq0$, corresponding to a topological phase with Chern number $C=1$ for $m<0$ and a trivial one with $C=0$ for $m>0$. In the original definition of the model~\cite{qi_prb2010} the SPE is accounted for by introducing a constant $s$-wave pairing $\Delta i\sigma_y$ in the BdG Hamiltonian, thus leaving the normal part unchanged. This approach, which we refer to as ``fixed-$\Delta$'', leads to a topological phase diagram of the SC with three different phases characterized by Chern numbers $\mathcal N=0,1,2$, also in the presence of doping~\cite{qi_prb2010}. We report our own numerical simulations in the fixed-$\Delta$ approximation at half-filling in the left-hand panel of Fig.~\ref{fig:qhz_comparison}. Here, the BdG Chern number $\mathcal N$ counts the number of chiral Majorana edge states crossing the gap of the BdG spectrum, and the phases with $\mathcal N=2$ and $\mathcal N=0$ are adiabatically connected respectively to the QAHI $C=1$ and $C=0$ topological phases. On the contrary, the $\mathcal N=1$ phase is a topological phase with no QAHI counterpart that is characterized by a single chiral Majorana edge mode.

        Although such minimal model provides a simple and useful platform to study both analytically and numerically the emergence of topological superconductivity in proximitized system, the fixed-$\Delta$ BdG approach might oversimplify the description of the SPE, thus hindering understanding. Here, we reformulate the QHZ model as separate QAHI and 2D $s$-wave SC in tight-binding, which are coupled by an interlayer hopping $\Gamma$. The heterostructure is studied through our embedding approach, Eqs.~\ref{eq:gel_embedding}--\ref{eq:induced_anomalous}, and we focus on the renormalization of the QAHI Green's function and its corresponding Chern number computed via the topological Hamiltonian method~\cite{wang_prx2012}. In this framework, an effective Hamiltonian is defined $H_{t}(\mathbf k)=-G_{\mathbf k}^{-1}(\omega=0)$ which allows calculating the topological invariant using the standard procedures (such as the plaquette method or the tracking of hermaphrodite Wannier function centers, a.k.a. Wilson loops~\cite{soluyanov_prb2011, gresch_prb2017, Vanderbilt2018}) for non-interacting systems even in a Green's functions framework. As in the original proposal~\cite{qi_prb2010}, the QAHI is represented by the QHZ model (Eq.~\ref{eq:qhz_hamilton}), here discretized on a square lattice, with parameters $A=B=1$. The $s$-wave SC is also on the square lattice and is described by the pairing amplitude $\Delta$. The interface hopping $\Gamma$ is diagonal in both spin and orbital space.
        \begin{figure*}
            \centering
            \includegraphics[width=\textwidth]{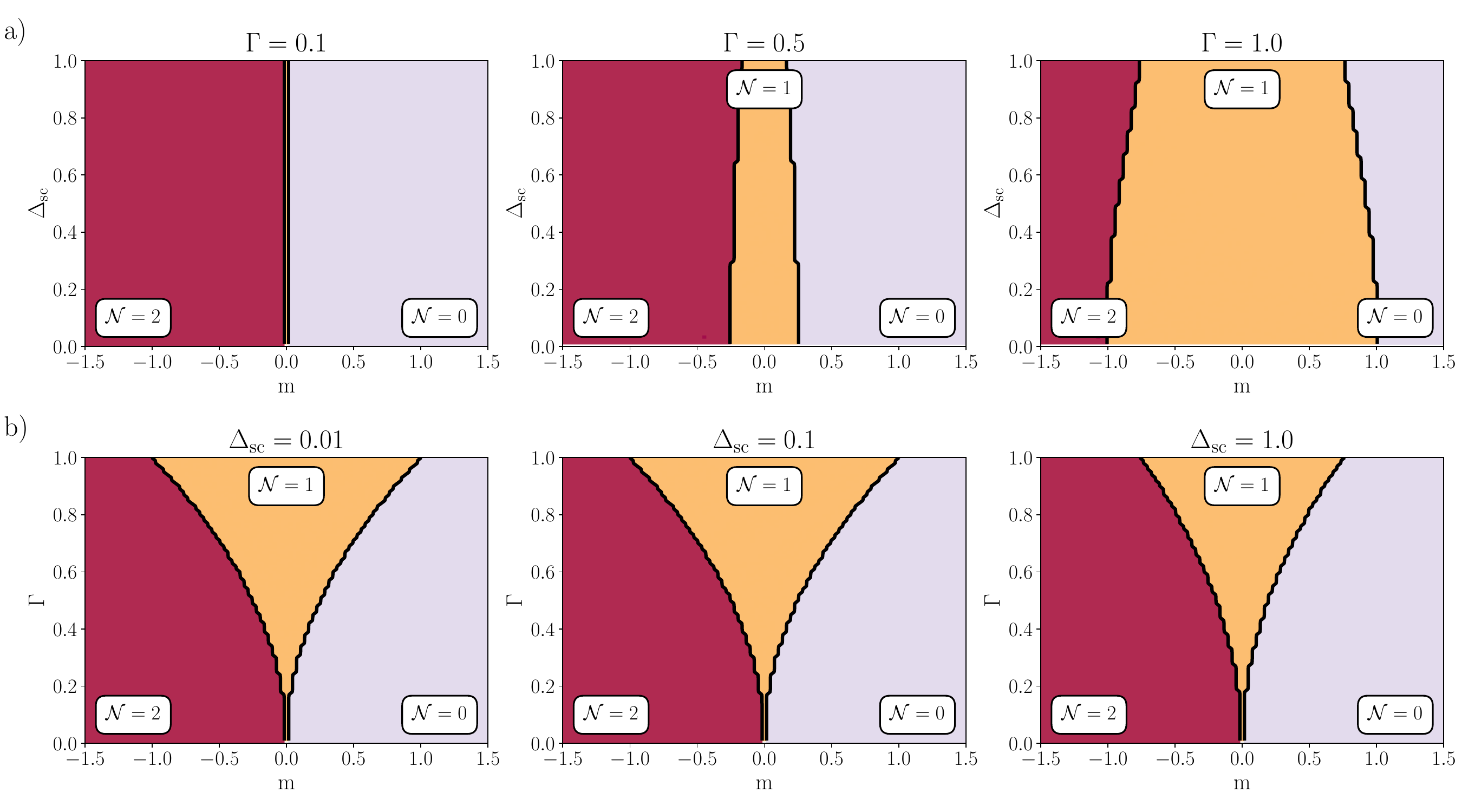}
            \caption{Topological phase diagrams obtained by solving the embedding equations for the QAHI/SC interface, model as in Fig.~\ref{fig:qhz_comparison}. In the top row (a), the value of the interface hopping $\Gamma$ is fixed at three different values (left, center and right panels) while the phase diagram is studied with respect to the bulk pairing $\Delta_{\mathrm{sc}}$. Vice versa, in the second row (b), $\Delta_{\mathrm{sc}}$ is fixed and the phase diagram is shown as function of $\Gamma$. The chiral topological phase with $\mathcal{N}=1$ becomes more stable for large values of the interlayer hopping. Notably, the $m\!-\!\Gamma$ phase diagram qualitatively resembles that obtained with the fixed-$\Delta$ approach (Fig.~\ref{fig:qhz_embedding}), where the dependence on $\Delta$ is hence better understood as a dependence on the interface coupling\textemdash which drives the induced anomalous density\textemdash rather than on the intrinsic pairing of the bulk SC.}
            \label{fig:qhz_embedding}
        \end{figure*}

        By solving the embedding equations at fixed interface coupling $\Gamma$ while varying the parent-SC pairing $\Delta_{\mathrm{sc}}$, the resulting topological phase diagram changes substantially relative to the conventional fixed-$\Delta$ treatment (see Fig.~\ref{fig:qhz_comparison}). Here, it should be noted that the $\Delta\rightarrow0$ limit has a different physical meaning in the two approaches. In the fixed-$\Delta$ case, this limit corresponds to a vanishing pairing that leaves the QAHI Hamiltonian unrenormalized\textemdash thus resulting in a BdG Chern number $\mathcal N$ that is twice that of the non-superconducting system $C$. By contrast, in the embedding approach, the interface hopping $\Gamma$ is finite, and the QAHI is still renormalized by the normal embedding potential $\nu_{\mathrm{no}}^{\mathrm{emb}}(\omega)$, which does not vanish when $\Delta_{\mathrm{sc}}\rightarrow0$. In addition, in this limit the topological invariant becomes ill-defined since the system becomes gapless: it is precisely the role of the pairing to open a gap in the spectrum, thereby allowing the definition of a topological invariant.

        To find an approximate limit in which the embedding equations (Eqs.~\ref{eq:gel_embedding}--\ref{eq:induced_anomalous}) resemble a fixed-$\Delta$ scheme, we can retain only the anomalous propagator of the SC, \(F_0^{(B)}\), while neglecting the normal (renormalizing) part of its embedding potential. To demonstrate this explicitly, we solve the embedding equations derived in Sec.~\ref{sec:ab_initio_spe} after dropping the second term in Eq.~\ref{eq:gel_embedding}, i.e. $G_{ne}^{(A)}(\omega)=G^{(A)}_0(\omega)$, which amounts to omitting the normal (renormalizing) contribution of the superconducting environment. Here, by varying the superconducting pairing $\Delta_{\mathrm{sc}}$, we can see that the fixed-$\Delta$ approach is similar to the approximation of the embedding equations in which $G_{ne}^{(A)}(\omega)=G^{(A)}_0(\omega)$, especially for small values of $\Delta_{\mathrm{sc}}$, while the full embedding equations give a different picture. An ``effective pairing'' in this limit can be derived analytically, with calculations reported in Appendix~\ref{app:effective_pairing}. We conclude the first renormalization of the electronic Green's function $G_{ne}^{(A)}(\omega)$ is crucial to be predictive when dealing with real heterostructures, as noted also in Refs.~\cite{miller_pr1968, stanescu_prb2010, sau_prb2010, potter_prb2011}. A more detailed analysis of the embedding equations reveals that, while the normal embedding potential $\nu_{\mathrm{no}}^{\mathrm{emb}}(\omega)\sim\Gamma^2$ (independent of the superconducting gap) the anomalous one follows $\nu_{\mathrm{an}}^{\mathrm{emb}}(\omega)\sim\Delta\Gamma^2$. Then, to recover the fixed-$\Delta$ limit from our embedding formalism, we would need $\Gamma^2\ll1$ while maintaining $\Delta\Gamma^2$ finite, resulting in $\Delta\gg1$\textemdash a regime that is experimentally unrealistic. Additionally, the two approaches differ in their limit for $\Delta\rightarrow0$: in the fixed-$\Delta$ case the Green's function of system $A$ remains completely unrenormalized, while in the embedding equations the normal component is still included, highlighting the fact that the SPE involves both normal and anomalous renormalizations, and not merely the latter. More importantly, the fixed-$\Delta$ approach does not allow to disentangle the contributions of the interface hopping $\Gamma$ from the superconducting pairing $\Delta_{\mathrm{sc}}$ of the parent SC, as they are effectively absorbed into a single parameter $\Delta$.

        Now we investigate how the topological phase diagram changes by varying the two relevant materials parameters responsible for the SPE, namely the interface hopping $\Gamma$ and the pairing in the parent SC $\Delta_{\mathrm{sc}}$ (see Fig.~\ref{fig:qhz_embedding}). Similarly to the fixed-$\Delta$ case, we see the emergence of three topologically different phases. We find that the $m\!-\!\Delta_{\mathrm{sc}}$ phase diagram closely resembles the full-embedding results of Fig.~\ref{fig:qhz_comparison}: increasing the interface hopping broadens the topological region with Chern number $\mathcal N=1$, especially at small values $\Delta_{\mathrm{sc}}$. Conversely, when $\Delta_{\mathrm{sc}}$ is kept constant and $\Gamma$ is varied, the resulting phase diagram qualitatively reproduces the fixed-$\Delta$ behavior displayed in Fig.~\ref{fig:qhz_comparison}. The  $m\!-\!\Gamma$ phase diagram does not depend strongly on $\Delta_{\mathrm{sc}}$, indicating that the fixed-$\Delta$ phase diagram originally discussed in Ref.~\cite{qi_prb2010} describes more the effect of the interface coupling rather than changing the pairing of the bath SC. The analysis demonstrates that the embedding formalism provides a transparent, conceptually simple yet powerful framework for disentangling the contributions to the SPE. By treating the normal and anomalous renormalizations separately, one can quantify their individual roles and systematically control their influence on the properties of the proximitized system.

    \subsection{Mesoscopic systems: metals and topological insulators proximitized by \texorpdfstring{$s$}{}-wave superconductors}\label{sec:semiinfinite}
        The SPE typically occurs at the interface of 3D materials, where its spatial extent can be large. Thanks to the recursive schemes discussed in Sec.~\ref{sec:LS}, we can apply our embedding approach to mesoscopic system, avoiding finite-size effects. In this framework, the spatial decay of the induced superconducting correlations and of other key properties can be analyzed explicitly as a function of the distance from the interface, provided the principal layer is chosen sufficiently large.

        We start by considering a single-band free-electron-like metal on the cubic lattice, in contact with a lattice-matched $s$-wave SC. The chemical potential of both systems is aligned in the middle of the band, while the nearest neighbor hopping $t$ has the same amplitude but opposite sign. The bulk superconducting pairing amplitude is set at $\Delta=0.01t$ and the nearest neighbor interface hopping is $\Gamma=0.1t$, diagonal in the spin indices. We can then solve the iterative equations with a sufficiently thick principal layer and then obtain the layer-dependent order parameter by taking the trace of the final anomalous Green's function over the orbitals centered in the sublayer of interest. Namely, we define
        \begin{align}\label{eq:paulidecomposition}
            \mathcal F_{r_{\!\perp}, \sigma\sigma'}=\mathrm{Tr}\left[\mathcal P_{r_{\!\perp}}F_{\sigma\sigma'}\mathcal P_{r_{\!\perp}}\right]
        \end{align}
        where $\sigma$, $\sigma'$ are spin indices and $\mathcal P_{r_{\!\perp}}$ is projector onto the MLWFs localized at distance $r_{\!\perp}$ from the interface. Equation~\ref{eq:paulidecomposition} is then decomposed on the basis of Pauli matrices $(\mathbb I, \boldsymbol\sigma)$
        \begin{align}
            \mathcal F_{r_{\!\perp}}=i\left[\mathrm f_{r_{\!\perp},0}\,\mathbb I+\boldsymbol{\mathrm{f}}_{r_{\!\perp}}\cdot\boldsymbol\sigma\right]\sigma_y
        \end{align}
        allowing us to define the singlet and triplet average order parameters as
        \begin{align}
            \psi_s(r_{\!\perp})&=\frac{1}{\beta N_{\mathbf k}}\sum_{n\mathbf k}|\mathrm{f}_{r_{\!\perp},0}(\mathbf k, i\omega_n)|, \\
            \psi_t(r_{\!\perp})&=\frac{1}{\beta N_{\mathbf k}}\sum_{n\mathbf k}|\boldsymbol{\mathrm{f}}_{r_{\!\perp}}(\mathbf k, i\omega_n)|
        \end{align}
        where $\omega_n=(2n+1)\pi/\beta$ are the fermionic Matsubara frequencies and $\beta$ is the inverse temperature of the system.

\begin{figure}
            \centering
            \includegraphics[width=\linewidth]{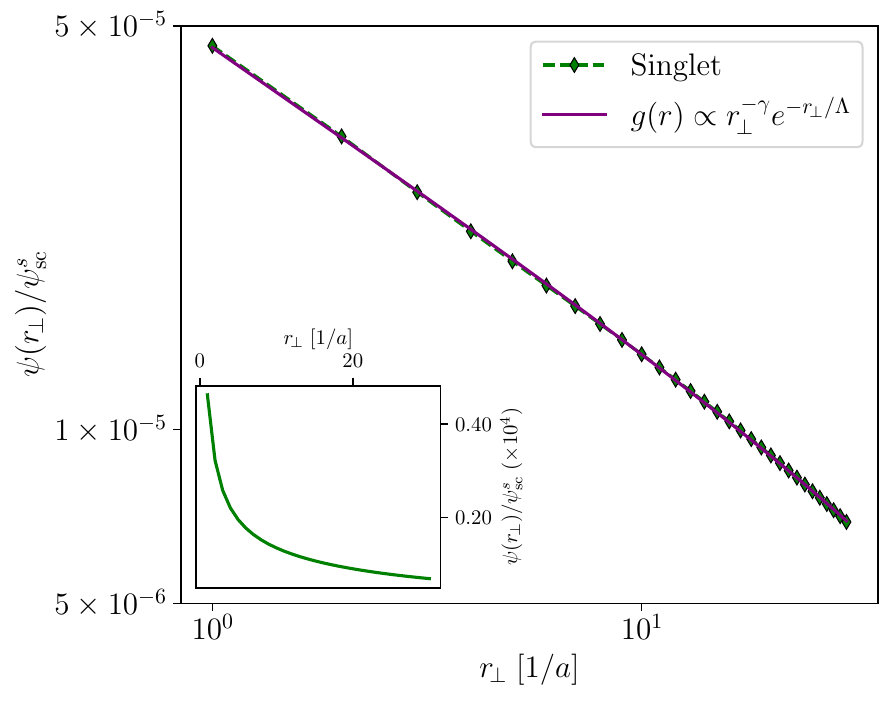}
            \caption{
            Layer-resolved singlet anomalous density induced in a semi-infinite, single-band, free-electron-like metal (cubic lattice) by contact with a semi-infinite \(s\)-wave superconductor. The induced anomalous density is plotted as a function of the distance from the interface \(r_{\!\perp}\) (measured in units of the lattice constant \(a\)).
            All data are normalized to the bulk \(s\)-wave anomalous density of the isolated superconductor, \(\psi^{s}_{\mathrm{sc}}\).
            The two subsystems are coupled via nearest-neighbor hopping of amplitude \(\Gamma=0.1\,t\), and the superconducting gap is \(\Delta=0.01\,t\). }
            \label{fig:penetration_depth_freeeg}
        \end{figure}

        The results for the SC/metal interface at temperature $1/\beta=10^{-4}$ eV are shown in Fig.~\ref{fig:penetration_depth_freeeg}, where we fit the anomalous density decay with a function $g(r_{\!\perp})= \alpha r_{\!\perp}^{-\gamma}e^{-r_{\!\perp}/\Lambda}$ and $\Lambda$ is defined as the superconducting proximity length. The singlet is the only induced component and is characterized by $\gamma=0.512\pm0.002$ and $\Lambda=198\pm9$ in units of the lattice parameter. Despite the induced anomalous density at the interface of the renormalized metal being about $5\cdot10^{-5}$ the anomalous density of the parent SC, the proximity length is very large, extending for many layers inside the bulk of the normal system.

        \begin{figure*}
            \centering
            \includegraphics[width=\linewidth]{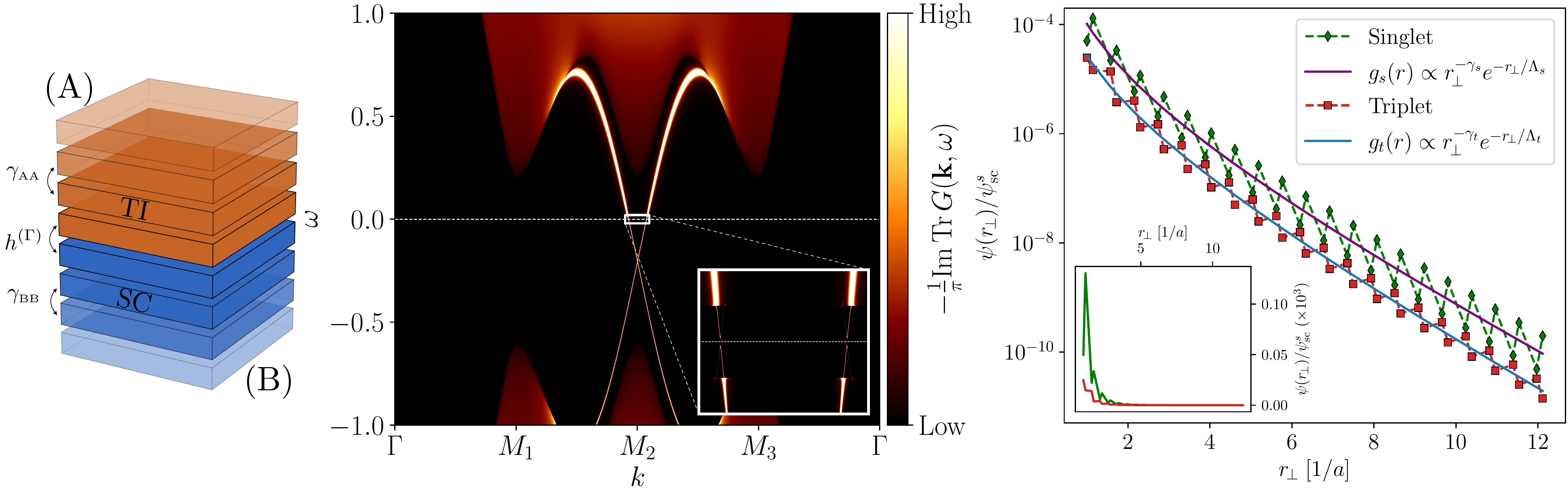}
            \caption{Left: Schematic of the interface between two semi-infinite subsystems \(A\) and \(B\), coupled by the interface hopping $h^{(\Gamma)}$. Each rectangular block denotes a principal layer in the López-Sancho iterative scheme; blocks of the same material are coupled to their nearest neighbors by the interlayer hopping \(\gamma\).
            Center: Renormalized spectral function of a Fu-Kane-Mele model in the $\mathbb{Z}_2$-strong \((1;000)\) topological phase brought into contact with a lattice-matched \(s\)-wave superconductor.  Right: Layer-resolved induced singlet and triplet anomalous densities plotted as a function of the distance from the interface \(r_{\!\perp}\) (measured in units of the lattice constant \(a\)). The anomalous densities are normalized to the bulk \(s\)-wave anomalous density of the isolated superconductor, \(\psi^{s}_{\mathrm{sc}}\). In all panels the two subsystems are coupled via nearest-neighbor interfacial hopping of amplitude \(\Gamma=0.1\,t\), and the superconducting pairing amplitude is \(\Delta=0.01\,t\).}
            \label{fig:renormalization_fkm}
        \end{figure*}
        Now we consider a different interface, where in place of the metal we study the tight-binding FKM model~\cite{fu_prl2007} on a diamond lattice, defined by the Hamiltonian
        \begin{align}
            \mathcal H_{\mathrm{FKM}}= \sum_{\langle ij\rangle}t_{ij}c_i^{\dagger}c_j+i\lambda_{\mathrm{so}}\!\sum_{\langle\!\langle ij\rangle\!\rangle}c_i^{\dagger}(\boldsymbol{\mathrm{g}}_{ij}\cdot\boldsymbol{\sigma})c_j + \mathrm{h.c.}
        \end{align}
        where $t_{ij}=t+\delta t_{ij}$ is the nearest neighbors hopping amplitude and the different values $\delta t_{ij}$ depend on the direction of the bond. In the second term, $\lambda_{\mathrm{so}}$ is the SOC strength, and $\boldsymbol{\mathrm{g}}_{ij}$ is a vector with length $\sqrt 2$ along the direction $\boldsymbol{\mathrm{d}}_{ik}\times\boldsymbol{\mathrm{d}}_{kj}$ where $\boldsymbol{\mathrm{d}}_{ik}$ and $\boldsymbol{\mathrm{d}}_{kj}$ are the nearest neighbors bond traversed to get from site $i$ to its next-nearest neighbor $j$. We then solve the embedding equations for the FKM model in the strong topological phase $(1;000)$ and coupled via nearest neighbors hopping to a lattice-matched $s$-wave SC, and we report the renormalized spectral function of the surface of the FKM model in Fig.~\ref{fig:renormalization_fkm}. The bulk superconducting pairing amplitude is set at $\Delta=0.01t$ and the nearest neighbor interface hopping is $\Gamma=0.1t$, diagonal in the spin indices. We set the chemical potential to $\mu=-0.2$ such that it lies in the bulk gap and above the surface Dirac point.

        By considering a thick principal layer, we can also compute the decay of the induced anomalous density as a function of the distance with respect to the interface. The results at temperature $1/\beta=10^{-4}$ eV are shown in Fig.~\ref{fig:renormalization_fkm}, where we can see that the average induced anomalous densities have the same behavior in the singlet and triplet channels. In particular, as previously, fits are performed with a function $g(r_{\!\perp})= \alpha r_{\!\perp}^{-\gamma}e^{-r_{\!\perp}/\Lambda}$ but for both components, yielding $\gamma_s=1.95\pm0.54$, $\Lambda_s=1.23\pm0.16$ for the singlet and $\gamma_t=1.81\pm0.30$, $\Lambda_t=1.15\pm0.08$ for the triplet. The induced anomalous density at the interface is about $10^{4}$ times smaller than the anomalous density in the bulk of the parent SC. Notably, the induced superconducting proximity length in this case is much smaller than the one obtained in the previous case for a metal. We interpret this difference as arising from the fact that the SPE here acts primarily on the edge states of the system, which are themselves exponentially localized at the surface, as opposed to the delocalized bulk metallic states of the single-band free-electron-like metal. Additionally, the presence of SOC allows for the emergence of a triplet component of the anomalous density, whose magnitude is about half the one of the singlet and has a similar decay behavior.

    \subsection{NbSe\texorpdfstring{$_2$}{}/CrBr\texorpdfstring{$_3$}{} heterostructure from first principles}\label{sec:scti}
        Having validated our embedding framework on model Hamiltonians, we now turn to a fully \textit{ab initio} application. Here, we investigate the SPE at the interface between the surface of the SC NbSe$_2$ and a monolayer of the 2D magnetic insulator CrBr$_3$. This heterostructure provides a realistic and technologically relevant test case, combining strong SOC, intrinsic magnetism, and multiband $s$-wave superconductivity~\cite{das_npj2023} within an interface in which the materials are weakly coupled via van der Waals interactions~\cite{kezilebieke_nature2020}.

        \begin{figure*}
            \centering
            \includegraphics[width=0.9\linewidth]{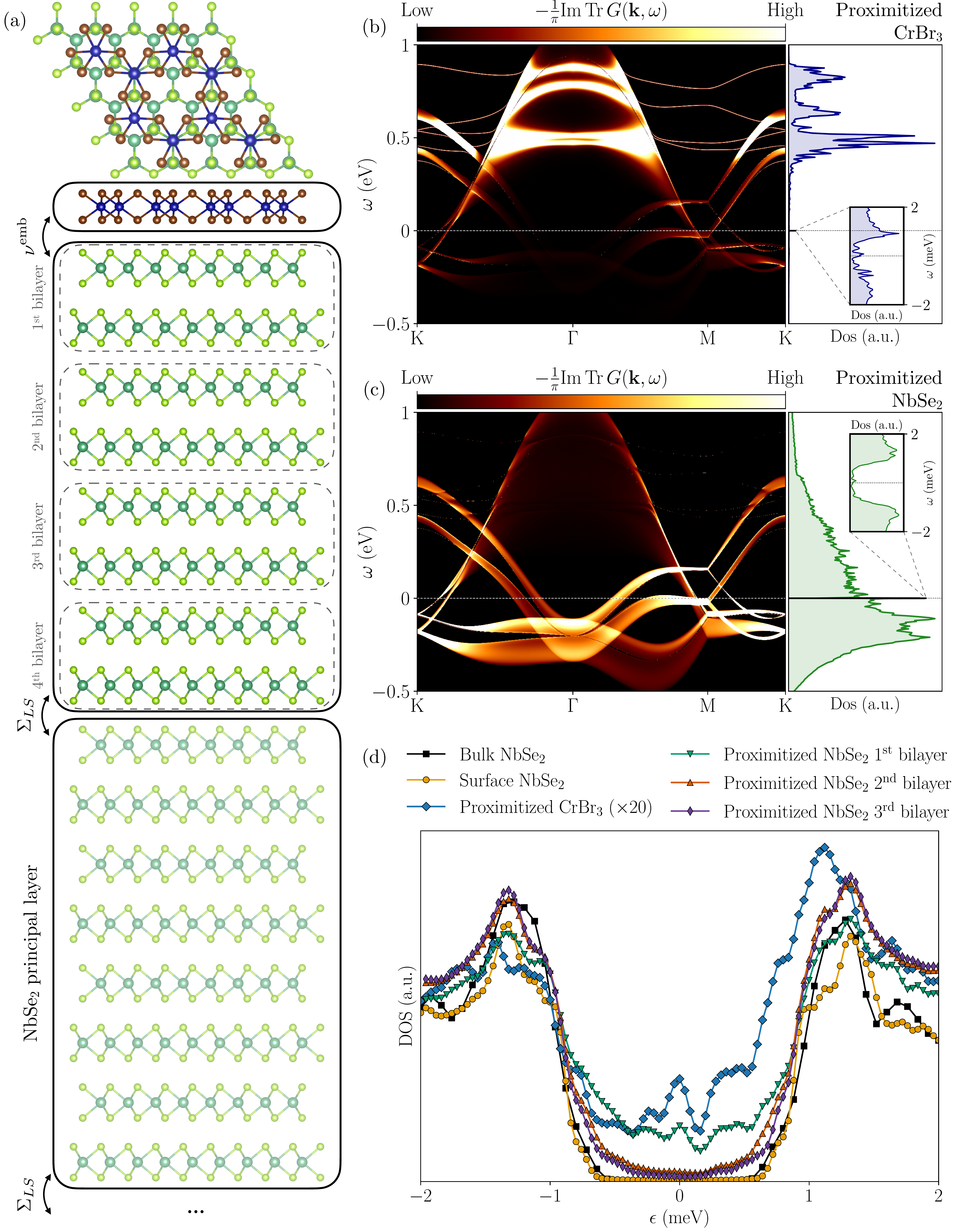}
            \caption{
            (a) Schematic of the NbSe$_2$/CrBr$_3$ heterostructure used in the embedding calculations. Top and side views of the interface are shown; solid boxes indicate the CrBr$_3$ layer and the principal layers of NbSe$_2$ (dashed boxes in the side view mark the sub-bilayers referred to in panel (d)).  (b) Renormalized spectral function and density of states (DOS) of CrBr$_3$ at the interface.  (c) Renormalized spectral function and DOS of the first principal layer of NbSe$_2$ at the interface.  (d) Comparison of DOS curves: pristine bulk NbSe$_2$, the surface DOS of isolated NbSe$_2$, the bilayer-resolved proximitized DOS at the interface (bilayers correspond to the dashed boxes in panel (a)), and the proximitized CrBr$_3$ DOS (rescaled for clarity).
            The proximity effect induces a very small but finite DOS at the Fermi level in CrBr$_3$ and causes a tiny reduction of the superconducting gap in NbSe$_2$, including surface in-gap states; nevertheless, the overall gap structure of NbSe$_2$ remains essentially preserved.}
            \label{fig:nbse2crbr3interface}
        \end{figure*}

        Experimentally, the inverse SPE on NbSe$_2$ appears to be relatively weak, yet its signatures remain subtle and somewhat debated. Several works observe a modest gap reduction and faint in-gap surface states~\cite{kezilebieke_nature2020, kezilebieke_adma2021, kezilebieke_nanolett2022}, whereas others find a DOS almost indistinguishable from pristine NbSe$_2$~\cite{li_natcomm2024, cuperus_scipost2025}. Importantly, there is currently no experimental consensus on the origin of the in-gap features; sample-dependent factors such as disorder, tip-sample coupling and scanning conditions, local stacking configuration at the interface, and relative twist angle between the layers can lead to partially different spectroscopic signatures and likely contribute to the spread of observations~\cite{kezilebieke_nature2020, kezilebieke_adma2021, kezilebieke_nanolett2022, li_natcomm2024, cuperus_scipost2025}.

        While here we do not aim at confirming or disproving individual experiments, we argue that our approach allows us to disentangle the intrinsic SPE from external factors, providing a clear picture of the expected signatures at the ideal interface and thus aiding the interpretation of experimental results. To achieve this, we perform first-principles DFT simulations of the isolated NbSe$_2$ and CrBr$_3$, as well as their interface, by using the Quantum ESPRESSO distribution~\cite{giannozzi_qe2009, giannozzi_qe2017,giann_jcp_2020} with norm-conserving fully relativistic pseudopotentials in the local spin-density approximation (LSDA)~\cite{vb_ssp_1972} obtained from the ONCVPSP code~\cite{oncvpsp_prb2013}. Then, we calculate from \textit{ab initio} the maximally-localized Wannier function (MLWF) Hamiltonian describing the Cr $d$ and Nb $d$ orbitals near the Fermi level using the Wannier90 code~\cite{marzari_prb_1997,Souza2001,Mostofi2008,Pizzi2020,marzari_rmp_2012, marrazzo_rmp_2024}. Details of the first-principles simulations of CrBr$_3$, NbSe$_2$ and their interface are reported in Appendix~\ref{app:nbse2crbr3}. Superconductivity in bulk NbSe$_2$ is modeled through two orbital-dependent $s$-wave pairings. These are first extracted from the values calculated \emph{ab initio} in Ref.~\cite{das_npj2023} for monolayer NbSe$_2$ near zero temperature, by using Migdal-Eliashberg MBPT and taking into account spin fluctuations. Then, these values are rescaled such that the critical temperature calculated by solving the BdG equations~\cite{zhu_bdgbook} for the bulk matches experimental results ($T_c = 7.2$ K). The interface hopping matrices are extracted directly from the Wannier Hamiltonian of the entire heterostructure, and we further approximate it by retaining only the hopping terms between CrBr$_3$ and the first NbSe$_2$ monolayer.

        With our embedding approach, we can study both the inverse SPE at the surface of a mesoscopic NbSe$_2$ substrate and the direct SPE on the CrBr$_3$ side of the heterostructure. In Fig.~\ref{fig:nbse2crbr3interface} we show the renormalized spectral function of CrBr$_3$ and of the NbSe$_2$ surface principal layer, the latter comprising $8$ NbSe$_2$ monolayers (in the following divided into 4 bilayers). As expected from the weak van der Waals coupling, the spectral density is dominated by the features of the two corresponding isolated systems; nevertheless, in both cases we observe an additional, weaker contribution that can be clearly attributed to proximity-induced hybridization with the other material.
        This is even more evident in the density of states (DOS) shown in the same figure: while the DOS of an isolated CrBr$_3$ monolayer vanishes at the Fermi level, the SPE induces a small but finite CrBr$_3$ DOS at the Fermi energy. Conversely, within a narrow energy window around the Fermi level, the NbSe$_2$ DOS is only weakly renormalized. The DOS are calculated over $300$ frequencies in a uniform $60\times60$ grid in reciprocal space, while the insets are calculated with a finer $120\times120$ grid and $100$ frequencies to better resolve the low-energy features.

        \begin{figure}
            \centering
            \includegraphics[width=\linewidth]{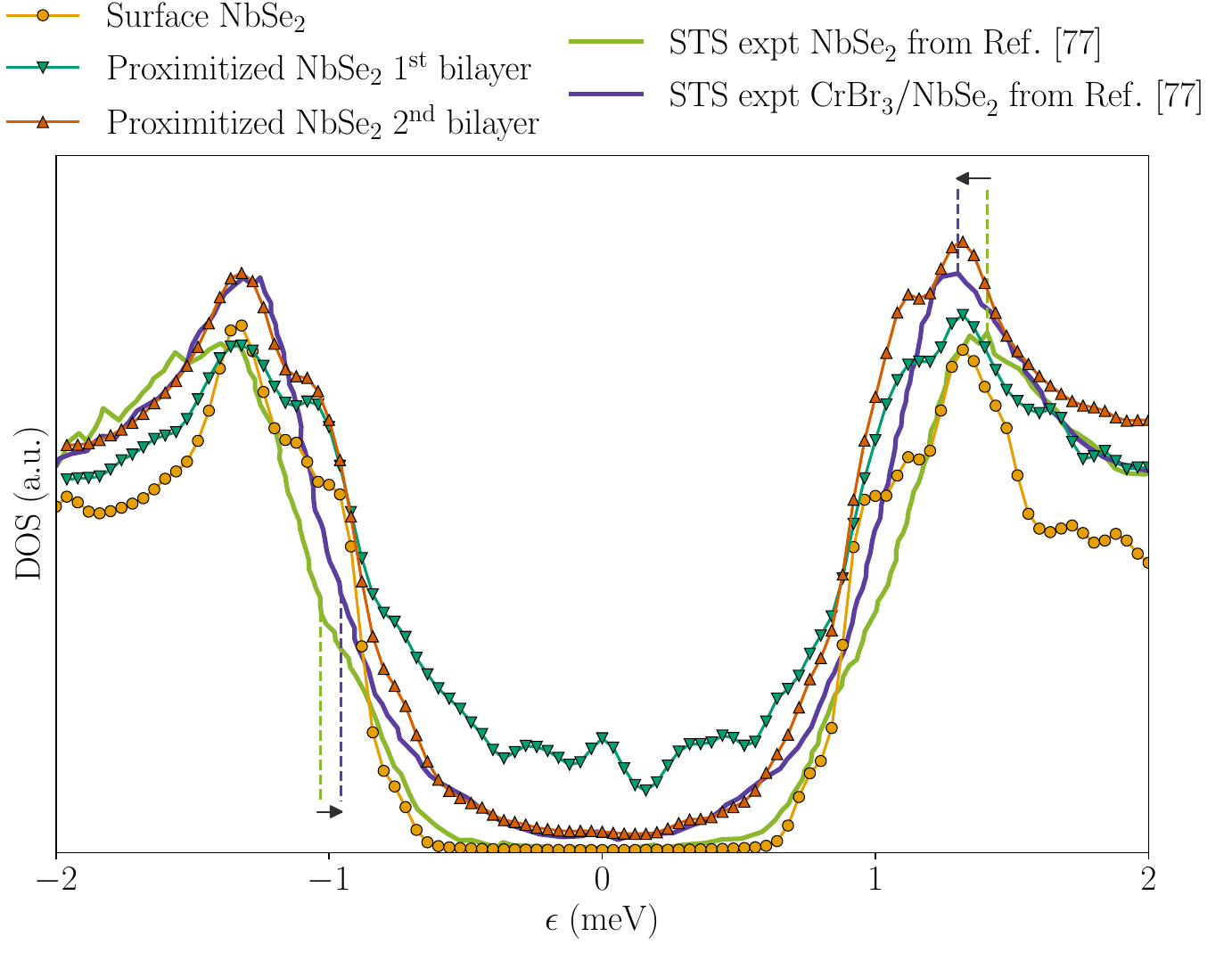}
            \caption{Experimental scanning tunneling spectroscopy (STS) data for bare NbSe$_2$ and the NbSe$_2$/CrBr$_3$ van der Waals heterostructure reported in Ref.~\cite{kezilebieke_nature2020}, compared with theoretical results obtained using our embedding approach [Eqs.~\ref{eq:gel_embedding}--\ref{eq:induced_anomalous}]. Our calculation qualitatively captures the two main experimental signatures: a slight reduction of the superconducting gap and the emergence of in-gap states in the heterostructure.}
            \label{fig:comparison_exp}
        \end{figure}
        To gain further insight, we focus on the low-energy DOS (bottom-right panel of Fig.~\ref{fig:nbse2crbr3interface}). The superconducting gap at the interface is slightly reduced compared with pristine NbSe$_2$, either in the bulk or at a surface in contact with vacuum. Notably, in-gap states appear at the interface but decay into the substrate, becoming very weak by the third NbSe$_2$ bilayer. Finally, CrBr$_3$ displays a proximity-induced finite DOS at the Fermi level despite the isolated monolayer being gapped, although its spectral weight is more than an order of magnitude smaller than that of NbSe$_2$ at the same energy.

        In Fig.~\ref{fig:comparison_exp}, we compare our results with the experimental scanning tunneling spectroscopy (STS) data of Ref.~\cite{kezilebieke_nature2020}. Our calculation reproduces the two main qualitative features observed experimentally: the modest reduction of the superconducting gap and the appearance of in-gap states at the interface. Overall, our embedding approach captures the essential experimental phenomenology despite the simplified treatment of lattice mismatch, superconductivity in NbSe$_2$, and other sample-dependent effects.

        \begin{figure}
            \centering
            \includegraphics[width=\linewidth]{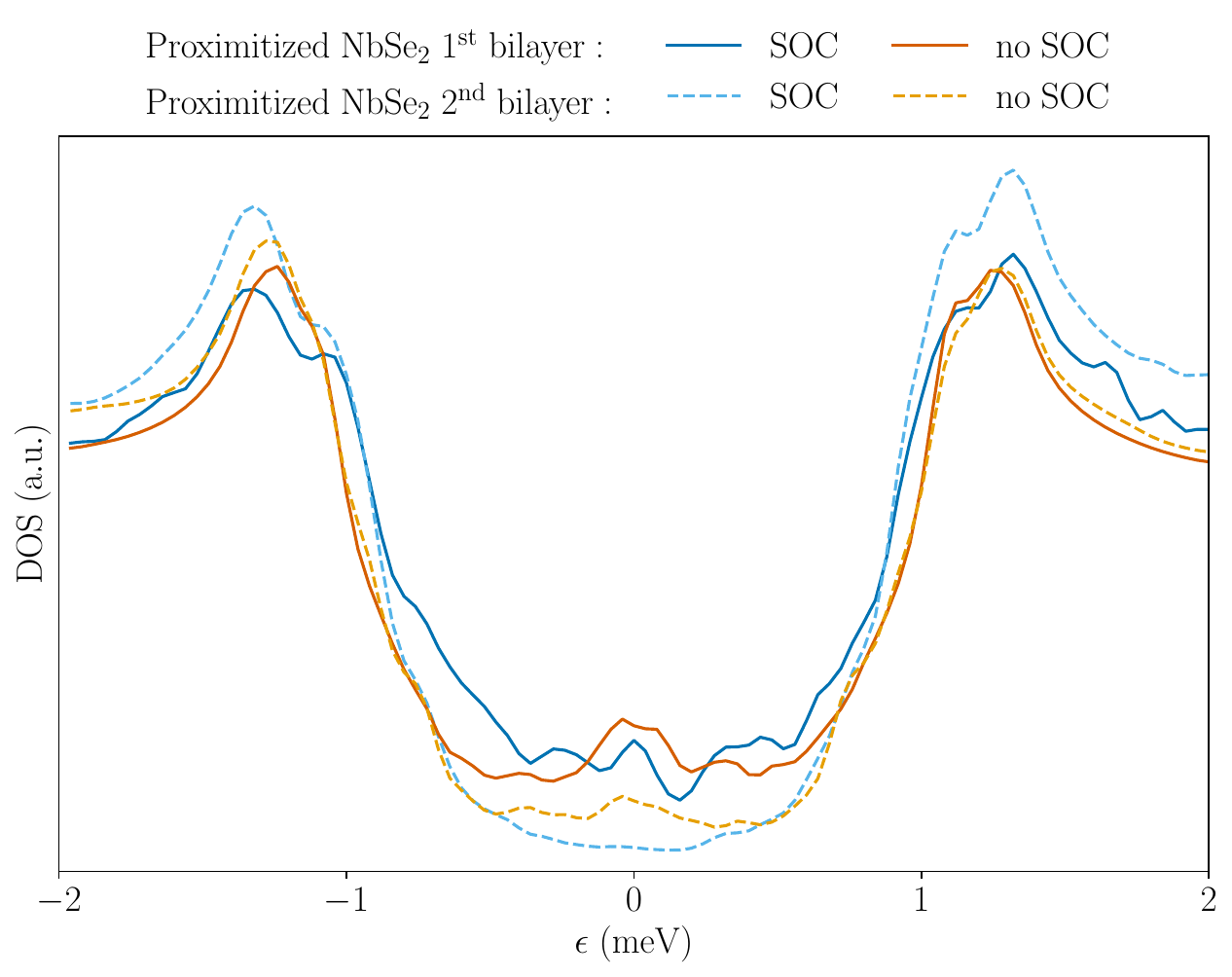}
            \caption{Calculated bilayer-resolved density of states (DOS) near the Fermi level for NbSe$_2$/CrBr$_3$ with and without spin-orbit coupling (SOC). The in-gap states persist even in the absence of SOC, showing that they originate from the magnetic character of CrBr$_3$ rather than from spin-orbit effects. In contrast, SOC produces a slight reduction of the superconducting gap, which is not observed when SOC is neglected. The in-gap states also decay more slowly into NbSe$_2$ without SOC, as reflected by the larger spectral weight at the Fermi level in the second bilayer.}
            \label{fig:abinitio_nosoc}
        \end{figure}
        Beyond comparison with STS experiments, our embedding framework allows us to disentangle the physical mechanisms responsible for specific spectral features and thus to guide their interpretation. As an illustration, in Fig.~\ref{fig:abinitio_nosoc} we repeat the first-principles calculations without SOC and compare them with the SOC-included results shown previously. The in-gap states persist in the absence of SOC, indicating that they originate from the magnetic character of CrBr$_3$ rather than from spin-orbit effects. By contrast, without SOC the superconducting gap remains essentially unchanged, while the in-gap states penetrate more deeply into NbSe$_2$, as reflected in their slower decay away from the interface. Overall, our results suggest that the experimental observations are consistent with an inverse magnetic proximity effect as the origin of the in-gap states, with SOC possibly accounting for the small gap reduction reported in Refs.~\cite{kezilebieke_nature2020,kezilebieke_adma2021,kezilebieke_nanolett2022}.

        We note that our embedding approach can be readily extended to more realistic interface geometries, including twisted moiré heterostructures~\cite{kezilebieke_nanolett2022}, to incorporate more sophisticated models of superconductivity in NbSe$_2$~\cite{noat_prb_2015,mazin_prx_2020,das_npj2023,Roy_2025}, or to perform systematic numerical studies of the effect of the twist angle~\cite{cuperus_scipost2025}; these materials-oriented applications are left for future work.

\section{Conclusions}\label{sec:conclusions}
    In this work, we have developed a first-principles Green's-function framework for the SPE based on real-space dynamical embedding and MLWF. The embedding equations (Eqs.~\ref{eq:gel_embedding}--\ref{eq:induced_anomalous}) for the SPE\textemdash particularly when represented diagrammatically as in Fig.~\ref{fig:embedding_eqs}\textemdash offer a transparent interpretation of the different contributions to the proximity effect in terms of normal and anomalous embedding potentials that renormalize the corresponding closed (isolated) systems. This formulation naturally yields systematic, physically motivated approximations that both clarify the underlying physical contributions and make it possible to scale the computational scheme to the long (hundreds of nm) length scales and the small (meV) energy scales that characterize the SPE.

    The extensive numerical validations presented in this work\textemdash including a variety of tight-binding models, 2D and mesoscopic 3D systems, and \emph{ab initio} simulations\textemdash demonstrate the practical utility, conceptual clarity, and versatility of the embedding scheme for a broad range of applications. As the 2D QAHI/SC example shows, the embedding approach can clarify the physics of simplified models by exposing their limitations (for instance, the common neglect of the normal contributions to the SPE), justifying their built-in approximations, and providing a first-principles interpretation of their effective parameters. Through numerical studies of interfaces between a SC and either a free-electron metal or a topological insulator, we highlight the importance of recursive embedding schemes to eliminate finite-size artifacts and to accurately probe the spatial decay of proximity effects into the bulk.

    More broadly, our embedding approach allows for the seamless combination of different levels of theory within a single framework, where each subsystem can be treated with the most appropriate method.
    This includes \textit{ab initio} approaches such as DFT, SCDFT, and MBPT\textemdash such as the Migdal-Eliashberg theory for phonon-mediated superconductivity and GW for electron-electron interactions\textemdash and, where necessary, DMFT to treat strong correlations in either the superconducting or normal component, which allow studying superconductivity in the very strong coupling limit~\cite{ToschiPRB2005,Witt2024} and describing polaronic superconductivity~\cite{Sous2023}.
    For interfaces with reasonably weak coupling (as is typical of van der Waals heterostructures), a major practical advantage is that one need not simulate an impractically thick superconducting slab. In other words, the embedding scheme alleviates the computational cost that makes direct, large-scale SCDFT or MBPT calculations prohibitive for realistic interfaces. We emphasize that our embedding approach is not an alternative to SCDFT or MBPT; rather, it is a higher-level theoretical and computational framework that can incorporate and leverage SCDFT, Migdal-Eliashberg, DMFT, or other methods to provide an efficient and accurate first-principles description of the SPE, while providing a transparent physical interpretation that disentangles the relevant contributions.
    As demonstrated by our \emph{ab initio} simulations of the CrBr$_3$/NbSe$_2$ heterostructure\textemdash and by comparison with the corresponding STS experiments~\cite{kezilebieke_nature2020}\textemdash the combination of real-space embedding and MLWF enables predictive descriptions of relevant materials platforms, the calculation of key observables, and the provision of microscopic insight that assists the interpretation of experimental measurements, while naturally leveraging a broad ecosystem of electronic-structure codes~\cite{marrazzo_rmp_2024}.

    Realistic interfaces, especially non-van der Waals ones, can host defects and impurities that may profoundly affect the SPE~\cite{belzig_prb1996, hui_prb2015, zhu_bdgbook}. More generally, the present formalism enables the study of induced superconductivity in disordered systems, such as alloys or non-crystalline materials, where a fully first-principles treatment of the superconducting interface may not be feasible. In particular, the embedding approach provides a flexible framework for incorporating disorder effects on either side of the interface, ranging from mean-field self-energies to more computationally intensive and accurate \emph{ab initio} self-consistent methods. This paves the way for future studies of increasingly realistic superconducting interfaces beyond ideal surfaces, in which disorder, defects, and structural complexity can be systematically incorporated within the same formalism.

    A further direction toward increased complexity is the treatment of electron-electron interactions. In particular, our equations can straightforwardly accommodate the direct inclusion of DMFT Green's functions, enabling extensions beyond conventional superconductivity in different directions, including (i) the possibility to address strongly coupled superconducting states (reaching the Bose-Eistein or bipolaronic regimes); (ii) the inclusion of pronounced or large electronic correlations as in cuprates, iron-based materials, fullerides and nickelates, and (iii) even scenarios in which superconductivity emerges at the interface despite being absent in the bulk constituents of the heterostructure. To include these effects, one could systematically study multilayer cuprates or superconducting heterostructures, where it has been argued that an enhancement of superconductivity arises from proximity effects between a strong-coupling SC and a weak-coupling one~\cite{Berg2008,Mazza2021,Smit2025}.

    Some of these applications require the inclusion of dynamical renormalization of the interface, including the coupling between the two subsystems, which is not included in the present formalism. Such an extension is nontrivial, as it requires an explicit treatment of cross-interactions between the subsystems. Likewise, the limit of strong interfacial coupling\textemdash i.e., when Eq.~\ref{eq:weak_limit_def} no longer provides a good approximation\textemdash can also be addressed by deriving and solving a self-consistent set of coupled equations for the two subsystems.
    These self-consistent and strongly-coupled extensions will be the subject of future work.

\section{Acknowledgements}
    A.M. acknowledges partial support from the PRIN Project ``Simultaneous electrical control of spin and valley polarisation in van der Waals magnetic materials'' (SECSY-CUP Grant No. J53D23001400001, PNRR Investimento M4.C2.1.1), M.D. acknowledges support from PNRR DM 118/2023 Investimento M4C1 3.4 T1, which are funded by the European Union - NextGenerationEU.
    M.C. acknowledges financial support from the National Recovery and Resilience Plan (PNRR) MUR PRIN 2022 Project (Prot. 20228YCYY7).
    The views and opinions expressed are solely those of the authors and do not necessarily reflect those of the European Union, nor can the European Union be held responsible for them. The authors acknowledge CINECA, under the ISCRA initiative and the CINECA-UniTS and CINECA-SISSA agreements, for the availability of high-performance computing resources and support, including simulation time on Galileo100 and Leonardo.

\section{Data Availability}
    The data that support the findings of this article are publicly available on the Materials Cloude Archive~\cite{materialscloud_data}.

\appendix
\section{Wannier functions and real-space partitioning scheme}\label{app:wannier}
    In Section~\ref{sec:ab_initio_spe} of the main text we introduce the embedding equations for the SPE in a fully general form, without specifying the representation in which the Green's function and the Hamiltonian are evaluated. In practice, however, the Green's function is non-local in real space, making a real-space solution of the embedding problem numerically demanding. Nonetheless, is possible to partition the reciprocal-space Hamiltonian by exploiting the exponential localization of the Wannier-function basis
    \begin{align}
        \mathcal H_{ij}^{(\alpha\beta)}(\mathbf{k})=\sum_{\mathbf{R}}\sum_{i\in\alpha,j\in\beta}e^{i\mathbf k\cdot\mathbf R}\mathcal H_{ij}(\mathbf{R}).
    \end{align}
    In this representation, the partitioning is entirely determined by the real-space support of the Wannier projectors, regardless of whether the underlying quantities are evaluated in real or reciprocal space. By the same token, the Green's function may be computed in reciprocal space and partitioned in the same manner. The locality of the Wannier basis thus yields a rigorous partitioning and an efficient route to implement the embedding formalism, without requiring an explicit real-space treatment of the Green's function.

\section{Inducing triplet superconductivity via the embedding equations}\label{app:triplet}
    SOC and magnetism are among the reported effects that can induce triplet superconductivity in a normal material by proximity effects~\cite{buzdin_rmp2005, bergeret_rmp2005, efetov_springer2008}. This can also be seen from the embedding framework we develop in the main text. We solve here explicitly Eq.~\ref{eq:induced_anomalous} for a normal material described by a single spinor in contact with an $s$-wave SC described a single spinor. For ease of notation, we rename the interface hopping $h^{(\Gamma)}=\Gamma$, and since all the quantities are evaluated at the same frequency, we drop the explicit frequency dependence, with the notation $G_{\alpha\beta}(\omega)=G_{\alpha\beta}$ and $G_{\alpha\beta}(-\omega)=\underline{G}_{\alpha\beta}$. Then, the induced anomalous Green's function in the normal material is obtained by solving
    \begin{widetext}\begin{align}\label{eq:explicit_triplet}
        F^{(A)\dagger}=-\begin{pmatrix}
            \underline{G}^{A}_{ne, \uparrow\uparrow} & \underline{G}^{A}_{ne, \uparrow\downarrow} \\
            \underline{G}^{A}_{ne, \downarrow\uparrow} & \underline{G}^{A}_{ne, \downarrow\downarrow}
        \end{pmatrix} \begin{pmatrix} \Gamma^*_{\uparrow\uparrow} & \Gamma^*_{\uparrow\downarrow} \\ \Gamma_{\uparrow\downarrow} & \Gamma^*_{\downarrow\downarrow} \end{pmatrix} \begin{pmatrix}
            0 & F^{(B)\dagger}_{0,\uparrow\downarrow} \\
            F^{(B)\dagger}_{0,\downarrow\uparrow} & 0
        \end{pmatrix} \begin{pmatrix} \Gamma^*_{\uparrow\uparrow} & \Gamma_{\uparrow\downarrow} \\ \Gamma^*_{\uparrow\downarrow} & \Gamma^*_{\downarrow\downarrow} \end{pmatrix} \begin{pmatrix}
            G^{(A)}_{\uparrow\uparrow} & G^{(A)}_{\uparrow\downarrow} \\
            G^{(A)}_{\downarrow\uparrow} & G^{(A)}_{\downarrow\downarrow}
        \end{pmatrix}
    \end{align}\end{widetext}
    which can be decomposed into a basis of Pauli matrices $(\mathbb I, \boldsymbol\sigma)$ as
    \begin{equation}
        F=i\left(\mathrm f_0\,\mathbb I+\boldsymbol{\mathrm{f}}\cdot\boldsymbol\sigma\right)\sigma_y,
    \end{equation}
    with $\mathbf{f}=(f_x,f_y,f_z)$. The components $f_{y/x} \propto F^{(A)\dagger}_{\uparrow\uparrow}\pm F^{(A)\dagger}_{\downarrow\downarrow}$ are zero if the interface hopping is diagonal in spin, i.e. $\Gamma_{\uparrow\downarrow}=0$, and the proximitized Green's functions $G^{(A)}$ and $G^{(A)}_{ne}$ of the normal material have no spin-flip components. Such spin-mixing contributions may arise from intrinsic SOC in the isolated materials, from proximity to a system with strong SOC, or from off-diagonal spin components in the interface hopping. The $f_z=\frac{1}{2}(F^{(A)\dagger}_{\uparrow\downarrow}+F^{(A)\dagger}_{\downarrow\uparrow})$ component, instead, can be nonzero even in the absence of SOC in the interface hopping and in the Green's functions of the normal material and the parent SC. In fact, the $f_z$ component is given by
    \begin{align}\begin{aligned}
        2f_z&= \underline{G}^{(A)}_{ne,\uparrow\uparrow}\Gamma^*_{\uparrow\uparrow}F^{(B)\dagger}_{0,\uparrow\downarrow}\Gamma^*_{\downarrow\downarrow}G^{(A)}_{\downarrow\downarrow}+ \\ &+G^{(A)}_{ne,\downarrow\downarrow}\Gamma^*_{\downarrow\downarrow}F^{(B)\dagger}_{0,\downarrow\uparrow}\Gamma^*_{\uparrow\uparrow}G^{(A)}_{\uparrow\uparrow},
    \end{aligned}\end{align}
    which vanishes with the additional constraint of time reversal symmetry. In this case, time reversal symmetry and the absence of SOC lead to $G_{\uparrow\uparrow}=G_{\downarrow\downarrow}$, and by using the additional particle-hole symmetry of the BdG formulation we conclude
    \begin{align}\begin{aligned}
        2f_z&= \underline{G}^{(A)}_{ne,\uparrow\uparrow}\Gamma^*_{\uparrow\uparrow}F^{(B)\dagger}_{0,\uparrow\downarrow}\Gamma^*_{\uparrow\uparrow}G^{(A)}_{\uparrow\uparrow}- \\ &-G^{(A)}_{ne,\uparrow\uparrow}\Gamma^*_{\uparrow\uparrow}F^{(B)\dagger}_{0,\uparrow\downarrow}\Gamma^*_{\uparrow\uparrow}G^{(A)}_{\uparrow\uparrow}=0.
    \end{aligned}\end{align}
    This shows that we can induce triplet superconductivity through the $f_z$ channel via a magnetic field even when the interface hopping is spin-diagonal and the normal material has no intrinsic SOC~\cite{buzdin_rmp2005}.

\section{Embedding equations in the strong interface coupling limit}\label{app:strongcoupling}
    In the strong coupling limit, three systems have to be considered in the embedding equations: the bulk normal material, the bulk SC and their interface, respectively regions $A$, $C$ and $B$ in the language of Fig.~\ref{fig:scheme}. This means that within this limit, in theories where the Hamiltonian is a functional of a global variable, the interface $B$ has to be converged self-consistently at the same level of theory. Then, after the projection, following Eq.~\ref{eq:projection}, we need to solve
    \begin{align}
        \begin{pmatrix}
                \omega-\mathcal H_{\mathrm{BdG}}^{(A)} & -\Gamma_{\mathrm{BdG}}^{(AB)} & 0 \\
                -\Gamma_{\mathrm{BdG}}^{(AB)\dagger} & \omega-\mathcal H_{\mathrm{BdG}}^{(B)} & -\Gamma_{\mathrm{BdG}}^{(BC)} \\
                0 & -\Gamma_{\mathrm{BdG}}^{(BC)\dagger} & \omega-\mathcal H_{\mathrm{BdG}}^{(C)}
            \end{pmatrix}\mathcal G(\omega)=\mathbb{I}.
    \end{align}
    Following a similar procedure as in the case of the embedding for two systems, we find that the normal Green's function of the interface is renormalized in three steps. First, we define the effective normal component of the Green's function of the interface
    \begin{align}\begin{aligned}\label{eq:tripleembedding_generic_intro}
        G_{e}^{(B)}(\omega)&=G_{0}^{(B)}(\omega)+\\ &+F_{0}^{(B)}(\omega)\left[G_{0}^{(B)T}(-\omega)\right]^{-1}F_{0}^{(B)\dagger}(\omega).
    \end{aligned}\end{align}
    Then, the two normal Green's function of systems $A$ and $C$ renormalize the effective normal Green's function at the interface
    \begin{align}\begin{aligned}
        G_{ne}^{(B)}& (\omega) =G_e^{(B)}(\omega)+\\ &+G_e^{(B)}(\omega)\left[\nu^{(A)}_{\mathrm{no}}(\omega)+\nu_{\mathrm{no}}^{(C)}(\omega)\right]G_{ne}^{(B)}(\omega)
    \end{aligned}\end{align}
    where the normal embedding potentials are defined as
    \begin{align}
        &\nu_{\mathrm{no}}^{(A)}(\omega)=\gamma^{(AB)\dagger}G_{0}^{(A)}(\omega)\gamma^{(AB)} \\
        &\nu_{\mathrm{no}}^{(C)}(\omega)=\gamma^{(BC)}G_{0}^{(C)}(\omega)\gamma^{(BC)\dagger}
    \end{align}
    with $\gamma^{(AB)}$ ($\gamma^{(BC)}$) the normal electronic component of the BdG hopping matrix $\Gamma^{(AB)}_{\mathrm{BdG}}$ ($\Gamma^{(BC)}_{\mathrm{BdG}}$).
    Then, $G_{ne}^{(B)}$ is renormalized by the anomalous component of the bulk SC $C$ as
    \begin{align}\begin{aligned}
        G^{(B)}(\omega)& =G_{ne}^{(B)}(\omega)+G_{ne}^{(B)}(\omega) \nu_{\mathrm{an}}^{(C)}(\omega) \cdot\\ &\cdot G_{ne}^{(B)T}(-\omega)\nu_{\mathrm{an}}^{(C)T}(-\omega)G^{(B)}(\omega)
    \end{aligned}\end{align}
    with the anomalous embedding potential
    \begin{align}\begin{aligned}\label{eq:anomalous_embedding_strong}
        \nu_{\mathrm{an}}^{(C)}(\omega)&=\left[G_{ne}^{(B)}(\omega)\right]^{-1}F_{0}^{(B)}(\omega)\left[G_{0}^{(B)T}(-\omega)\right]^{-1}+ \\ &+\gamma^{(BC)}F_{0}^{(C)}(\omega)\gamma^{(BC)T}.
    \end{aligned}\end{align}
    Interestingly, Eq.~\ref{eq:anomalous_embedding_strong} contains a self-embedding term that survives also when $\gamma_{AB},\gamma_{BC}\rightarrow 0$, and is responsible for the recovery of the isolated Green's function $G^{(B)}\rightarrow G_{0}^{(B)}$ in that limit. Finally, the induced anomalous Green's function reads
    \begin{align}\label{eq:tripleembedding_generic_final}
        F^{(B)\dagger}(\omega)=G_{ne}^{(B)T}(-\omega)\nu^{(C),T}_{\mathrm{an}}(-\omega)G^{(B)}(\omega).
    \end{align}
    With these equations we can recover the case of a normal system renormalized by a SC by setting $F_0^{(B)}=0$ and $\gamma^{(AB)}=0$, as in the main text. In addition, we have also access to other cases, for instance when a SC is renormalized by another SC, that can be obtained by setting $\gamma^{(AB)}=0$, or a SC renormalized by a normal material\textemdash i.e., the inverse proximity effect, as is the case for NbSe$_2$ renormalized by CrBr$_3$ discussed in Sec.~\ref{sec:scti}\textemdash that can be obtained by setting $\gamma^{(BC)}=0$. It is worth noting that this approach can be recursively scaled to study systems with an arbitrary number of partitions, in the spirit of Refs.~\cite{lopezsancho_jpf1984,lopezsancho_jpf1985}. If we divide a system into $N$ partitions coupled by nearest-partition hoppings only, we can compute the renormalized BdG Green's function of the $n$-th partition $\mathbb{G}^{(n)}(\omega)$ by solving
    \begin{align}\begin{aligned}
        \mathbb{G}^{(n)}&(\omega)= \mathbb{G}_0^{(n)}(\omega)+\\ &+\mathbb{G}_0^{(n)}(\omega)\left[\nu_+^{(n)}(\omega)+\nu_-^{(n)}(\omega)\right]\mathbb{G}^{(n)}(\omega),
    \end{aligned}\end{align}
    where the BdG embedding self-energies are defined through the recursive relations
    \begin{align}\begin{cases}
        \nu_-^{(n)}(\omega)=\Gamma_{n-1,n}^{\dagger}\tilde{\mathbb{G}}_{n-1}(\omega)\Gamma_{n-1,n} \\
        \tilde{\mathbb{G}}_{n-j}=\left[\mathbb{G}_0^{(n-j), -1}(\omega)-\nu_-^{(n-j)}(\omega)\right]^{-1}
    \end{cases}\end{align}
    and similarly
    \begin{align}\begin{cases}
        \nu_+^{(n)}(\omega)=\Gamma_{n,n+1}\tilde{\mathbb{G}}_{n+1}(\omega)\Gamma^{\dagger}_{n,n+1} \\
        \tilde{\mathbb{G}}_{n+j}=\left[\mathbb{G}_0^{(n+j), -1}(\omega)-\nu_+^{(n+j)(\omega)}\right]^{-1}
    \end{cases}\end{align}
    where $\Gamma_{ij}$ are the BdG hopping matrices between systems $i$ and $j$. Starting from the outermost partitions, $\tilde{\mathbb{G}}_{1}(\omega)=\mathbb{G}_0^{(1)}(\omega)=(\omega-\mathcal H_{\mathrm{BdG}}^{(1)})^{-1}$ and $\tilde{\mathbb{G}}_{N}(\omega)=\mathbb{G}_0^{(N)}(\omega)=(\omega-\mathcal H_{\mathrm{BdG}}^{(N)})^{-1}$, we can embed subsequent partitions by using the appropriate limit in the solution of the embedding equations for the general case (Eqs.~\ref{eq:tripleembedding_generic_intro}--\ref{eq:tripleembedding_generic_final}). In this way, we can solve the equations iteratively until we reach the desired partition $n$.\\

\section{Effective pairing from the purely anomalous renormalization of the embedding equations}\label{app:effective_pairing}
    In the main text we argue that the fixed-$\Delta$ approach can be interpreted as a purely anomalous embedding, in which the normal embedding contributions are neglected, that is $G_{ne}^{(A)}(\omega) \approx G_0^{(A)}(\omega)$. Here, we solve analytically the embedding equations for the triplet anomalous Green's function for the QHZ model in contact with an $s$-wave SC using this approximation. We report the triplet anomalous Green's function of the QHZ model, respectively as obtained with the fixed-$\Delta$ approach and by solving the embedding equations while neglecting $\nu_{\mathrm{no}}^{\mathrm{emb}}$:
    \begin{widetext}\begin{align}\label{eq:triplet_qhz_fixd}
       [\textrm{Fixed-}\Delta] \hspace{6pt} F^{\dagger}_{\uparrow\uparrow}(\mathbf k, \omega)=\frac{2\Delta\epsilon_{\mathbf k}A_{\mathbf k}^*}{\left[\omega^2-(\epsilon_{\mathbf k}-\Delta)^2-|A_{\mathbf k}|^2\right]\left[\omega^2-(\epsilon_{\mathbf k}+\Delta)^2-|A_{\mathbf k}|^2\right]}
    \end{align}\begin{align}\label{eq:triplet_emb_approx}
        [\nu_{\mathrm{no}}^{\mathrm{emb}}\!=\!0] \hspace{6pt} F^{\dagger}_{\uparrow\uparrow}(\mathbf k, \omega)=\frac{2\Gamma^2\Delta\epsilon_{\mathbf k}A_{\mathbf k}^*(\omega^2-\xi_{\mathbf k}^2-\Delta^2)^3}{(\omega^2-\epsilon_{\mathbf k}^2-|A_{\mathbf k}|^2)^2(\omega^2-\xi_{\mathbf k}^2-\Delta^2)^4-2\Gamma^4\Delta^2(\omega^2+\epsilon_{\mathbf k}^2-|A_{\mathbf k}|^2)(\omega^2-\xi_{\mathbf k}^2-\Delta^2)^2+\Gamma^8\Delta^4}
    \end{align}\end{widetext}
    where $\epsilon_{\mathbf k}$ and $A_{\mathbf k}$ are defined in Eq.~\ref{eq:qhz_hamilton} for the QHZ model, while $\xi_{\mathbf k}$ and $\Delta$ are the normal dispersion and the pairing potential of the parent $s$-wave SC. In particular, we can rewrite Eq.~\ref{eq:triplet_emb_approx} in a form similar to Eq.~\ref{eq:triplet_qhz_fixd}, identifying an ``effective induced pairing'' $\Delta_{\mathrm{eff}}$ having the same role of $\Delta$ in the fixed-$\Delta$ case. Note that in the proximitized system the pairing potential is always zero (only the anomalous density is non-vanishing) and $\Delta_{\mathrm{eff}}$ is just referred to as \textit{pairing} for the sake of comparison with the fixed-$\Delta$ case. If, in the fixed-$\Delta$ case, $\Delta\ll\epsilon_{\mathbf k}$, we can neglect terms of order $\Delta^4$ in the denominator of Eq.~\ref{eq:triplet_qhz_fixd}. Then, aside from a multiplicative function proportional to $\Gamma^2F_0^{(B)}(\mathbf k, \omega)$, Eq.~\ref{eq:triplet_emb_approx} is similar in the two approaches by defining the effective induced pairing $\Delta_{\mathrm{eff}}$ as
    \begin{widetext}\begin{align}
        \Delta_{\mathrm{eff}}(\mathbf k, \omega)=\Gamma^2F_0^{(B)}(\mathbf k, \omega)\sqrt{\dfrac{1}{2}\Bigg(1+\dfrac{\omega^2}{\epsilon_{\mathbf k}^2}-\dfrac{|A_{\mathbf k}|^2}{\epsilon_{\mathbf k}^2}\Bigg)-\Bigg(\dfrac{\Gamma^2F_0^{(B)}(\mathbf k, \omega)}{2\epsilon_{\mathbf k}}\Bigg)^2}.
    \end{align}\end{widetext}
    By further simplification, if $\Gamma\ll\epsilon_{\mathbf k}$ we can neglect the term proportional to $\Gamma^4$ in the square root, and we restrict the frequencies to those corresponding to bands in the normal material $\omega=E_{\mathbf k}$, with $E_{\mathbf k}=\sqrt{\epsilon_{\mathbf k}^2+|A_{\mathbf k}|^2}$. Then, the effective induced pairing in the limit where the normal embedding contributions are neglected is given by
    \begin{align}
        \Delta_{\mathrm{eff}}(\mathbf k)\approx\Gamma^2F_0^{(B)}(\mathbf k, E_{\mathbf k}).
    \end{align}
    This result is consistent with the common assumption that the effective induced pairing in the normal material is proportional to the anomalous Green's function of the parent SC, with a proportionality constant given by the square of the interface hopping $\Gamma^2$. However, we stress that this result is only valid in the limit where the normal embedding contributions are neglected, and it can be significantly modified when the full embedding equations are solved, as shown in the main text for the QHZ model.

\section{Details on first principles simulations of the NbSe\texorpdfstring{$_2$}{}/CrBr\texorpdfstring{$_3$}{} heterostructure}\label{app:nbse2crbr3}
\begin{figure*}
        \centering
        \includegraphics[width=0.8\linewidth]{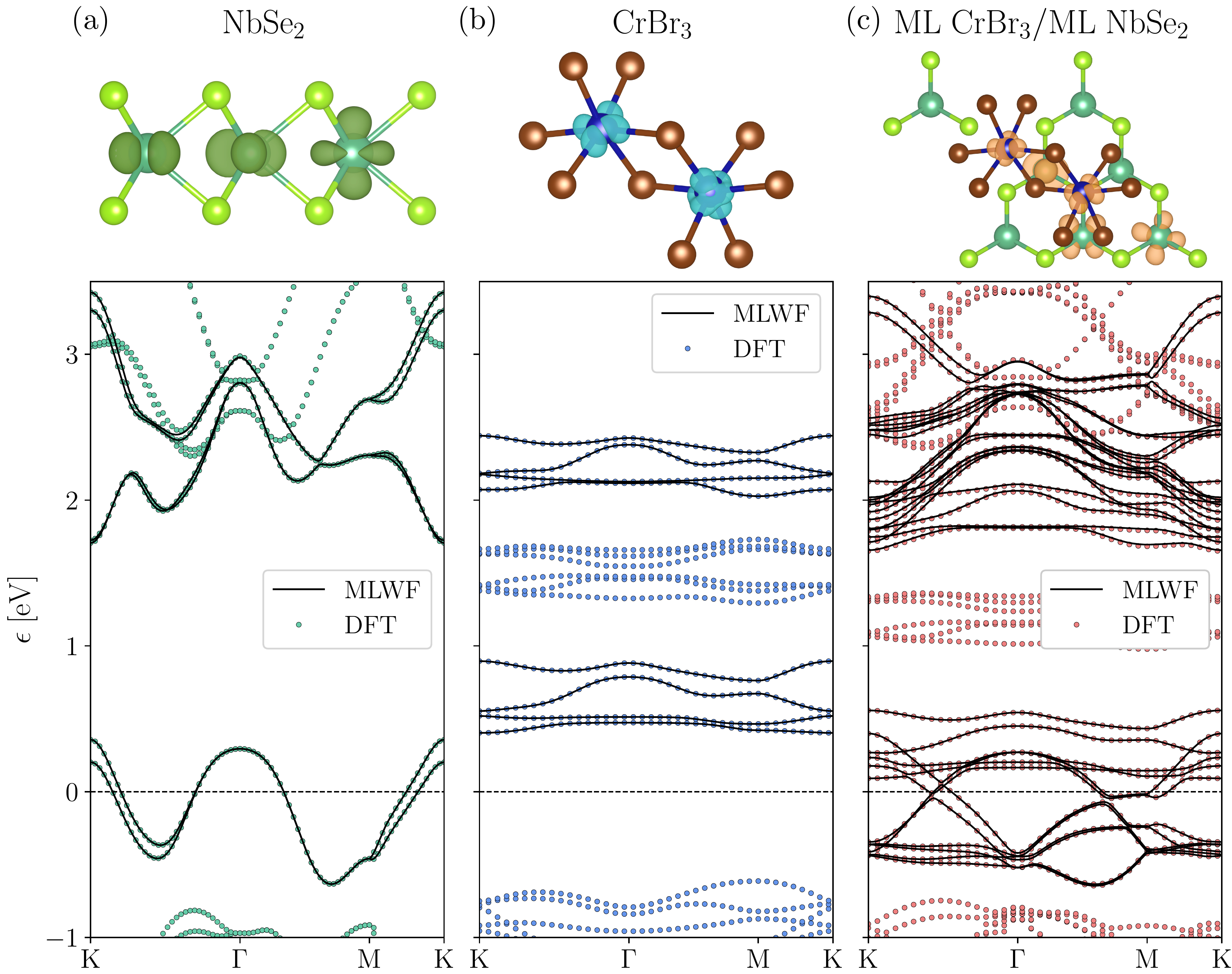}
        \caption{Maximally-localized Wannier functions (MLWFs) and comparison between the DFT and the Wannier-interpolated band structure for (a) isolated NbSe$_2$, (b) isolated CrBr$_3$ and (c) NbSe$_2$/CrBr$_3$ heterostructure. The Wannier Hamiltonian accurately reproduces the DFT band structure in all cases, while the MLWFs of the heterostructure match those of the isolated systems.}
        \label{fig:dftwannier}
    \end{figure*}

    First-principles DFT simulations have been performed using the Quantum ESPRESSO distribution~\cite{giannozzi_qe2009, giannozzi_qe2017, giann_jcp_2020}. Due to the lattice mismatch between CrBr$_3$ and NbSe$_2$, we model their interface as a unit cell of CrBr$_3$ stacked on top of a $2\times2$ supercell of NbSe$_2$. We simulate the htCrSe stacking, in which one Cr is directly above a Se site and the other is in a hollow site of NbSe$_2$ (see Fig.~\ref{fig:nbse2crbr3interface}), as this was reported to be most energetically stable stacking in the literature~\cite{kezilebieke_nature2020}. We perform variable-cell optimization of the lattice parameters and atomic positions. In this case, calculations are done without SOC, using norm-conserving PBE pseudopotentials from the PseudoDojo library~\cite{oncvpsp_prb2013, vansetten_cpc2018} for NbSe$_2$ and ultrasoft pseudopotentials from the SSSP PBE Precision 1.3.0 library~\cite{sssplib, garrity_cms2014} for CrBr$_3$. In structural optimization, we account for van der Waals interactions through the rVV10 functional~\cite{v_prl_2009,sab_prb_2013}. The final relaxed lattice shows that CrBr$_3$ is strained by about $7.8\%$ with respect to its isolated lattice parameter, while NbSe$_2$ is compressed by about $0.6\%$. We neglect the small compression in the simulation of isolated NbSe$_2$, while we use the relaxed lattice parameter at the interface to simulate isolated CrBr$_3$. In both cases, we used the LSDA functional and norm-conserving fully relativistic pseudopotentials obtained from the ONCVPSP code~\cite{oncvpsp_prb2013}. Then, we calculate the MLWF Hamiltonian describing the Cr $d$ and Nb $d$ orbitals near the Fermi level using the Wannier90 code~\cite{marzari_prb_1997,Souza2001,Mostofi2008,Pizzi2020,marrazzo_rmp_2024}. The initial projections are chosen to be atomic-like $d_{z^2}$, $d_{xy}$ and $d_{x^2-y^2}$ orbitals centered on the Nb atoms (as the major contribution to the bands around Fermi is due to these orbitals in the scalar relativistic approximation, as also noted by Ref.~\cite{Roy_2025}) and automated projections obtained through the SCDM method~\cite{damle2015, damle2017} on the Cr atoms. We target $6$ Wannier functions in the isolated NbSe$_2$ and $8$ in the isolated CrBr$_3$, resulting in $32$ Wannier functions at the interface between the monolayers. We find that the resulting Wannier Hamiltonian accurately reproduces the DFT band structure, as reported in Fig.~\ref{fig:dftwannier}. The matching between the MLWF at the interface and the corresponding Wannier functions of the isolated materials is done by comparing the spin and real-space character of each Wannier function. The embedding equations in the main text are then solved by using the Wannier Hamiltonian of the isolated systems coupled by the interface hopping extracted from the interface calculation.

\bibliography{biblio}

@PREAMBLE{
 "\providecommand{\noopsort}[1]{}" 
 # "\providecommand{\singleletter}[1]{#1}%" 
}

@article{lee_prl_2005,
  title = {{Band Structure and Quantum Conductance of Nanostructures from Maximally Localized Wannier Functions: The Case of Functionalized Carbon Nanotubes}},
  author = {Lee, Young-Su and Nardelli, Marco Buongiorno and Marzari, Nicola},
  journal = {Phys. Rev. Lett.},
  volume = {95},
  issue = {7},
  pages = {076804},
  numpages = {4},
  year = {2005},
  month = {Aug},
  publisher = {American Physical Society},
  doi = {10.1103/PhysRevLett.95.076804},
}

@article{nardelli_prb_1999,
  title = {Electronic transport in extended systems: {Application} to carbon nanotubes},
  author = {Nardelli, Marco Buongiorno},
  journal = {Phys. Rev. B},
  volume = {60},
  issue = {11},
  pages = {7828--7833},
  numpages = {0},
  year = {1999},
  month = {Sep},
  publisher = {American Physical Society},
  doi = {10.1103/PhysRevB.60.7828},
}

@article{marrazzo_rmp_2024,
  title = {Wannier-function software ecosystem for materials simulations},
  author = {Marrazzo, Antimo and Beck, Sophie and Margine, Elena R. and Marzari, Nicola and Mostofi, Arash A. and Qiao, Junfeng and Souza, Ivo and Tsirkin, Stepan S. and Yates, Jonathan R. and Pizzi, Giovanni},
  journal = {Rev. Mod. Phys.},
  volume = {96},
  issue = {4},
  pages = {045008},
  numpages = {54},
  year = {2024},
  month = {Dec},
  publisher = {American Physical Society},
  doi = {10.1103/RevModPhys.96.045008},
}

@article{marzari_rmp_2012,
  title = {{Maximally localized Wannier functions: Theory and applications}},
  author = {Marzari, Nicola and Mostofi, Arash A. and Yates, Jonathan R. and Souza, Ivo and Vanderbilt, David},
  journal = {Rev. Mod. Phys.},
  volume = {84},
  issue = {4},
  pages = {1419--1475},
  numpages = {0},
  year = {2012},
  month = {Oct},
  publisher = {American Physical Society},
  doi = {10.1103/RevModPhys.84.1419},
}

@article{giustino_rmp_2017,
  title = {Electron-phonon interactions from first principles},
  author = {Giustino, Feliciano},
  journal = {Rev. Mod. Phys.},
  volume = {89},
  issue = {1},
  pages = {015003},
  numpages = {63},
  year = {2017},
  month = {Feb},
  publisher = {American Physical Society},
  doi = {10.1103/RevModPhys.89.015003},
}

@article{margine_prb_2013,
  title = {Anisotropic Migdal-Eliashberg theory using Wannier functions},
  author = {Margine, E. R. and Giustino, F.},
  journal = {Phys. Rev. B},
  volume = {87},
  issue = {2},
  pages = {024505},
  numpages = {12},
  year = {2013},
  month = {Jan},
  publisher = {American Physical Society},
  doi = {10.1103/PhysRevB.87.024505},
}

@article{yazdani_science2023,
  author = {Ali Yazdani  and Felix von Oppen  and Bertrand I. Halperin  and Amir Yacoby },
  title = {{Hunting for Majoranas}},
  journal = {Science},
  volume = {380},
  number = {6651},
  pages = {eade0850},
  year = {2023},
  doi = {10.1126/science.ade0850}
}

@article{flototto_sciadv2018,
    author = {David Fl\"{o}totto  and Yuichi Ota  and Yang Bai  and Can Zhang  and Kozo Okazaki  and Akihiro Tsuzuki  and Takahiro Hashimoto  and James N. Eckstein  and Shik Shin  and Tai-Chang Chiang },
    title = {Superconducting pairing of topological surface states in bismuth selenide films on niobium},
    journal = {Science Advances},
    volume = {4},
    number = {4},
    pages = {eaar7214},
    year = {2018},
    doi = {10.1126/sciadv.aar7214}
}

@article{mandal_cm2023,
    author = {Mandal, Manasi and Drucker, Nathan C. and Siriviboon, Phum and Nguyen, Thanh and Boonkird, Artittaya and Lamichhane, Tej Nath and Okabe, Ryotaro and Chotrattanapituk, Abhijatmedhi and Li, Mingda},
    title = {{Topological Superconductors from a Materials Perspective}},
    journal = {Chemistry of Materials},
    volume = {35},
    number = {16},
    pages = {6184-6200},
    year = {2023},
    doi = {10.1021/acs.chemmater.3c00713},
}

@article{fu_prl2008,
  title = {{Superconducting Proximity Effect and Majorana Fermions at the Surface of a Topological Insulator}},
  author = {Fu, Liang and Kane, C. L.},
  journal = {Phys. Rev. Lett.},
  volume = {100},
  issue = {9},
  pages = {096407},
  numpages = {4},
  year = {2008},
  month = {Mar},
  publisher = {American Physical Society},
  doi = {10.1103/PhysRevLett.100.096407},
}

@article{qi_prb2010,
  title = {{Chiral topological superconductor from the quantum Hall state}},
  author = {Qi, Xiao-Liang and Hughes, Taylor L. and Zhang, Shou-Cheng},
  journal = {Phys. Rev. B},
  volume = {82},
  issue = {18},
  pages = {184516},
  numpages = {5},
  year = {2010},
  month = {Nov},
  publisher = {American Physical Society},
  doi = {10.1103/PhysRevB.82.184516},
}

@article{lutchyn_prl2010,
  title = {{Majorana Fermions and a Topological Phase Transition in Semiconductor-Superconductor Heterostructures}},
  author = {Lutchyn, Roman M. and Sau, Jay D. and Das Sarma, S.},
  journal = {Phys. Rev. Lett.},
  volume = {105},
  issue = {7},
  pages = {077001},
  numpages = {4},
  year = {2010},
  month = {Aug},
  publisher = {American Physical Society},
  doi = {10.1103/PhysRevLett.105.077001},
}

@article{volkov_physica1995,
  title = {{Proximity and Josephson effects in superconductor-two-dimensional electron gas planar junctions}},
  journal = {Physica C: Superconductivity},
  volume = {242},
  number = {3},
  pages = {261-266},
  year = {1995},
  issn = {0921-4534},
  doi = {https://doi.org/10.1016/0921-4534(94)02429-4},
  author = {A.F. Volkov and P.H.C. Magnée and B.J. {van Wees} and T.M. Klapwijk}
}

@Article{hu_npj2025,
  author={Hu, Jingnan and Luo, Aiyun and Wang, Zhijun and Zou, Jingyu and Wu, Quansheng and Xu, Gang},
  title={A numerical method for designing topological superconductivity induced by s-wave pairing},
  journal={npj Computational Materials},
  year={2025},
  month={May},
  day={14},
  volume={11},
  number={1},
  pages={133},
  issn={2057-3960},
  doi={10.1038/s41524-025-01621-6},
}

@article{russmann_prr2023,
  title = {{Interorbital Cooper pairing at finite energies in Rashba surface states}},
  author = {R\"u\ss{}mann, Philipp and Bahari, Masoud and Bl\"ugel, Stefan and Trauzettel, Bj\"orn},
  journal = {Phys. Rev. Res.},
  volume = {5},
  issue = {4},
  pages = {043181},
  numpages = {17},
  year = {2023},
  month = {Nov},
  publisher = {American Physical Society},
  doi = {10.1103/PhysRevResearch.5.043181},
}

@article{luders_prb2005,
  title = {{Ab initio theory of superconductivity. I. Density functional formalism and approximate functionals}},
  author = {L\"uders, M. and Marques, M. A. L. and Lathiotakis, N. N. and Floris, A. and Profeta, G. and Fast, L. and Continenza, A. and Massidda, S. and Gross, E. K. U.},
  journal = {Phys. Rev. B},
  volume = {72},
  issue = {2},
  pages = {024545},
  numpages = {17},
  year = {2005},
  month = {Jul},
  publisher = {American Physical Society},
  doi = {10.1103/PhysRevB.72.024545},
}

@article{marques_prb2005,
  title = {{Ab initio theory of superconductivity. II. Application to elemental metals}},
  author = {Marques, M. A. L. and L\"uders, M. and Lathiotakis, N. N. and Profeta, G. and Floris, A. and Fast, L. and Continenza, A. and Gross, E. K. U. and Massidda, S.},
  journal = {Phys. Rev. B},
  volume = {72},
  issue = {2},
  pages = {024546},
  numpages = {13},
  year = {2005},
  month = {Jul},
  publisher = {American Physical Society},
  doi = {10.1103/PhysRevB.72.024546},
}

@article{linscheid_prb2015A,
  title = {{Ab initio theory of superconductivity in a magnetic field. I. Spin density functional theory for superconductors and Eliashberg equations}},
  author = {Linscheid, A. and Sanna, A. and Essenberger, F. and Gross, E. K. U.},
  journal = {Phys. Rev. B},
  volume = {92},
  issue = {2},
  pages = {024505},
  numpages = {19},
  year = {2015},
  month = {Jul},
  publisher = {American Physical Society},
  doi = {10.1103/PhysRevB.92.024505},
}

@article{linscheid_prb2015B,
  title = {{Ab initio theory of superconductivity in a magnetic field. II. Numerical solution}},
  author = {Linscheid, A. and Sanna, A. and Gross, E. K. U.},
  journal = {Phys. Rev. B},
  volume = {92},
  issue = {2},
  pages = {024506},
  numpages = {15},
  year = {2015},
  month = {Jul},
  publisher = {American Physical Society},
  doi = {10.1103/PhysRevB.92.024506},
}

@article{reho_prb2024,
  title = {{Density functional Bogoliubov-de Gennes theory for superconductors implemented in the SIESTA code}},
  author = {Reho, R. and Wittemeier, N. and Kole, A. H. and Ordej\'on, P. and Zanolli, Z.},
  journal = {Phys. Rev. B},
  volume = {110},
  issue = {13},
  pages = {134505},
  numpages = {19},
  year = {2024},
  month = {Oct},
  publisher = {American Physical Society},
  doi = {10.1103/PhysRevB.110.134505},
}

@article{reho_prb2026,
  title = {{Proximity-induced superconductivity in PbTe/Pb heterostructures from first principles}},
  author = {Reho, R. and Botello-M\'endez, A. R. and Zanolli, Zeila},
  journal = {Phys. Rev. B},
  volume = {113},
  issue = {5},
  pages = {054501},
  numpages = {11},
  year = {2026},
  month = {Feb},
  publisher = {American Physical Society},
  doi = {10.1103/r366-j5k6},
}

@article{lopezsancho_jpf1985,
doi = {10.1088/0305-4608/15/4/009},
year = {1985},
month = {apr},
publisher = {},
volume = {15},
number = {4},
pages = {851},
author = {M. P. Lopez Sancho and J. M. Lopez Sancho and J. M. L. Sancho and J. Rubio},
title = {{Highly convergent schemes for the calculation of bulk and surface Green functions}},
journal = {Journal of Physics F: Metal Physics}
}

@article{lopezsancho_jpf1984,
  doi = {10.1088/0305-4608/14/5/016},
  year = {1984},
  month = {may},
  publisher = {},
  volume = {14},
  number = {5},
  pages = {1205},
  author = {M. P. Lopez Sancho and J. M. Lopez Sancho and J. Rubio},
  title = {{Quick iterative scheme for the calculation of transfer matrices: application to Mo (100)}},
  journal = {Journal of Physics F: Metal Physics}
}

@book{tkachov_quantummaterials,
  author = {G. Tkachov},
  title = {{Topological Quantum Materials: Concepts, Models, and Phenomena}},
  publisher = {Jenny Stanford Publishing},
  year = {2022},
  doi = {10.1201/9781003266419}
}

@article{tkachov_prb2013,
  title = {Suppression of surface $p$-wave superconductivity in disordered topological insulators},
  author = {Tkachov, G.},
  journal = {Phys. Rev. B},
  volume = {87},
  issue = {24},
  pages = {245422},
  numpages = {7},
  year = {2013},
  month = {Jun},
  publisher = {American Physical Society},
  doi = {10.1103/PhysRevB.87.245422},
}

@article{meissner_1932,
  author={Holm, R. and Meissner, W.},
  title={{M}essungen mit {H}ilfe von fl{\"u}ssigem {H}elium. {XIII}},
  journal={{Z}eitschrift f{\"u}r {P}hysik},
  year={1932},
  month={Nov},
  day={01},
  volume={74},
  number={11},
  pages={715-735},
  issn={0044-3328},
  doi={10.1007/BF01340420},
}

@article{andreev_1964,
  author       = {Andreev, A F},
  title        = {THERMAL CONDUCTIVITY OF THE INTERMEDIATE STATE OF SUPERCONDUCTORS},
  journal      = {Zh. Eksperim. i Teor. Fiz.  },
  volume       = {Vol: 46},
  place        = {Country unknown/Code not available},
  year         = {1964},
  month        = {05}
}

@article{majorana_1937,
  author={Majorana, Ettore},
  title={Teoria simmetrica dell'elettrone e del positrone},
  journal={{Il Nuovo Cimento} (1924-1942)},
  year={1937},
  month={Apr},
  day={01},
  volume={14},
  number={4},
  pages={171-184},
  issn={1827-6121},
  doi={10.1007/BF02961314},
}

@article{linder_2015,
  author={Linder, Jacob and Robinson, Jason W. A.},
  title={Superconducting spintronics},
  journal={Nature Physics},
  year={2015},
  month={Apr},
  day={01},
  volume={11},
  number={4},
  pages={307-315},
  issn={1745-2481},
  doi={10.1038/nphys3242},
}

@article{linder_rmp2019,
  title = {Odd-frequency superconductivity},
  author = {Linder, Jacob and Balatsky, Alexander V.},
  journal = {Rev. Mod. Phys.},
  volume = {91},
  issue = {4},
  pages = {045005},
  numpages = {56},
  year = {2019},
  month = {Dec},
  publisher = {American Physical Society},
  doi = {10.1103/RevModPhys.91.045005},
}

@article{golubov_rmp2004,
  title = {The current-phase relation in Josephson junctions},
  author = {Golubov, A. A. and Kupriyanov, M. Yu. and Il'ichev, E.},
  journal = {Rev. Mod. Phys.},
  volume = {76},
  issue = {2},
  pages = {411--469},
  numpages = {0},
  year = {2004},
  month = {Apr},
  publisher = {American Physical Society},
  doi = {10.1103/RevModPhys.76.411},
}

@article{qi_rmp2011,
  title = {Topological insulators and superconductors},
  author = {Qi, Xiao-Liang and Zhang, Shou-Cheng},
  journal = {Rev. Mod. Phys.},
  volume = {83},
  issue = {4},
  pages = {1057--1110},
  numpages = {0},
  year = {2011},
  month = {Oct},
  publisher = {American Physical Society},
  doi = {10.1103/RevModPhys.83.1057},
}

@article{alicea_2012,
  doi = {10.1088/0034-4885/75/7/076501},
  year = {2012},
  month = {jun},
  publisher = {IOP Publishing},
  volume = {75},
  number = {7},
  pages = {076501},
  author = {Alicea, Jason},
  title = {{New directions in the pursuit of Majorana fermions in solid state systems}},
  journal = {Reports on Progress in Physics}
}

@article{kitaev_2001,
  doi = {10.1070/1063-7869/44/10S/S29},
  year = {2001},
  month = {oct},
  publisher = {},
  volume = {44},
  number = {10S},
  pages = {131},
  author = {A Yu Kitaev},
  title = {{Unpaired Majorana fermions in quantum wires}},
  journal = {Physics-Uspekhi}
}

@article{sarma_npj2015,
  author={Sarma, Sankar Das and Freedman, Michael and Nayak, Chetan},
  title={Majorana zero modes and topological quantum computation},
  journal={npj Quantum Information},
  year={2015},
  month={Oct},
  day={27},
  volume={1},
  number={1},
  pages={15001},
  issn={2056-6387},
  doi={10.1038/npjqi.2015.1},
}

@misc{russmann_arxiv2022,
  title={Proximity induced superconductivity in a topological insulator}, 
  author={Philipp R{\"u}{\ss}mann and Stefan Bl{\"u}gel},
  year={2022},
  eprint={2208.14289},
  archivePrefix={arXiv},
  primaryClass={cond-mat.mes-hall},
}

@article{stanescu_prb2010,
  title = {Proximity effect at the superconductor--topological insulator interface},
  author = {Stanescu, Tudor D. and Sau, Jay D. and Lutchyn, Roman M. and Das Sarma, S.},
  journal = {Phys. Rev. B},
  volume = {81},
  issue = {24},
  pages = {241310},
  numpages = {4},
  year = {2010},
  month = {Jun},
  publisher = {American Physical Society},
  doi = {10.1103/PhysRevB.81.241310},
}

@article{sau_prb2010,
  title = {{Robustness of Majorana fermions in proximity-induced superconductors}},
  author = {Sau, Jay D. and Lutchyn, Roman M. and Tewari, Sumanta and Das Sarma, S.},
  journal = {Phys. Rev. B},
  volume = {82},
  issue = {9},
  pages = {094522},
  numpages = {7},
  year = {2010},
  month = {Sep},
  publisher = {American Physical Society},
  doi = {10.1103/PhysRevB.82.094522},
}

@article{potter_prb2011,
  title = {Engineering a $p+\mathit{ip}$ superconductor: Comparison of topological insulator and Rashba spin-orbit-coupled materials},
  author = {Potter, Andrew C. and Lee, Patrick A.},
  journal = {Phys. Rev. B},
  volume = {83},
  issue = {18},
  pages = {184520},
  numpages = {11},
  year = {2011},
  month = {May},
  publisher = {American Physical Society},
  doi = {10.1103/PhysRevB.83.184520},
}

@article{miller_pr1968,
  title = {{Theory of Superconductor---Normal-Metal Interfaces}},
  author = {McMillan, W. L.},
  journal = {Phys. Rev.},
  volume = {175},
  issue = {2},
  pages = {559--568},
  numpages = {0},
  year = {1968},
  month = {Nov},
  publisher = {American Physical Society},
  doi = {10.1103/PhysRev.175.559},
}

@article{russmann_prb2022,
  title = {{Density functional Bogoliubov-de Gennes analysis of superconducting Nb and Nb(110) surfaces}},
  author = {R\"u\ss{}mann, Philipp and Bl\"ugel, Stefan},
  journal = {Phys. Rev. B},
  volume = {105},
  issue = {12},
  pages = {125143},
  numpages = {11},
  year = {2022},
  month = {Mar},
  publisher = {American Physical Society},
  doi = {10.1103/PhysRevB.105.125143},
}

@article{suvasini_prb1993,
  title = {Computational aspects of density-functional theories of superconductors},
  author = {Suvasini, M. B. and Temmerman, W. M. and Gyorffy, B. L.},
  journal = {Phys. Rev. B},
  volume = {48},
  issue = {2},
  pages = {1202--1210},
  numpages = {0},
  year = {1993},
  month = {Jul},
  publisher = {American Physical Society},
  doi = {10.1103/PhysRevB.48.1202},
}

@Article{silvert_1975,
  author={Silvert, William},
  title={Theory of the superconducting proximity effect},
  journal={Journal of Low Temperature Physics},
  year={1975},
  month={Sep},
  day={01},
  volume={20},
  number={5},
  pages={439-477},
  issn={1573-7357},
  doi={10.1007/BF00120864},
}

@article{petocchi_prb2016,
  title = {Embedding dynamical mean-field theory for superconductivity in layered materials and heterostructures},
  author = {Petocchi, Francesco and Capone, Massimo},
  journal = {Phys. Rev. B},
  volume = {93},
  issue = {23},
  pages = {235125},
  numpages = {9},
  year = {2016},
  month = {Jun},
  publisher = {American Physical Society},
  doi = {10.1103/PhysRevB.93.235125},
}

@article{fu_prl2007,
  title = {{Topological Insulators in Three Dimensions}},
  author = {Fu, Liang and Kane, C. L. and Mele, E. J.},
  journal = {Phys. Rev. Lett.},
  volume = {98},
  issue = {10},
  pages = {106803},
  numpages = {4},
  year = {2007},
  month = {Mar},
  publisher = {American Physical Society},
  doi = {10.1103/PhysRevLett.98.106803},
}

@article{volpez_prl2019,
  title = {{Second-Order Topological Superconductivity} in $\ensuremath{\pi}$-{Junction Rashba Layers}},
  author = {Volpez, Yanick and Loss, Daniel and Klinovaja, Jelena},
  journal = {Phys. Rev. Lett.},
  volume = {122},
  issue = {12},
  pages = {126402},
  numpages = {6},
  year = {2019},
  month = {Mar},
  publisher = {American Physical Society},
  doi = {10.1103/PhysRevLett.122.126402},
}

@article{zhu_prl2019,
  title = {{Second-Order Topological Superconductors with Mixed Pairing}},
  author = {Zhu, Xiaoyu},
  journal = {Phys. Rev. Lett.},
  volume = {122},
  issue = {23},
  pages = {236401},
  numpages = {8},
  year = {2019},
  month = {Jun},
  publisher = {American Physical Society},
  doi = {10.1103/PhysRevLett.122.236401},
}

@article{read_prb2000,
  title = {{Paired states of fermions in two dimensions with breaking of parity and time-reversal symmetries and the fractional quantum Hall effect}},
  author = {Read, N. and Green, Dmitry},
  journal = {Phys. Rev. B},
  volume = {61},
  issue = {15},
  pages = {10267--10297},
  numpages = {0},
  year = {2000},
  month = {Apr},
  publisher = {American Physical Society},
  doi = {10.1103/PhysRevB.61.10267},
}

@article{sato_rpp2017,
  doi = {10.1088/1361-6633/aa6ac7},
  year = {2017},
  month = {may},
  publisher = {IOP Publishing},
  volume = {80},
  number = {7},
  pages = {076501},
  author = {Sato, Masatoshi and Ando, Yoichi},
  title = {Topological superconductors: a review},
  journal = {Reports on Progress in Physics}
}

@article{khalaf_prb2018,
  title = {Higher-order topological insulators and superconductors protected by inversion symmetry},
  author = {Khalaf, Eslam},
  journal = {Phys. Rev. B},
  volume = {97},
  issue = {20},
  pages = {205136},
  numpages = {13},
  year = {2018},
  month = {May},
  publisher = {American Physical Society},
  doi = {10.1103/PhysRevB.97.205136},
}

@article{zhu_prb2018,
  title = {{Tunable Majorana corner states in a two-dimensional second-order topological superconductor induced by magnetic fields}},
  author = {Zhu, Xiaoyu},
  journal = {Phys. Rev. B},
  volume = {97},
  issue = {20},
  pages = {205134},
  numpages = {7},
  year = {2018},
  month = {May},
  publisher = {American Physical Society},
  doi = {10.1103/PhysRevB.97.205134},
}

@article{geier_prb2018,
  title = {Second-order topological insulators and superconductors with an order-two crystalline symmetry},
  author = {Geier, Max and Trifunovic, Luka and Hoskam, Max and Brouwer, Piet W.},
  journal = {Phys. Rev. B},
  volume = {97},
  issue = {20},
  pages = {205135},
  numpages = {33},
  year = {2018},
  month = {May},
  publisher = {American Physical Society},
  doi = {10.1103/PhysRevB.97.205135},
}

@article{yang_jpcm2024,
  doi = {10.1088/1361-648X/ad3abd},
  year = {2024},
  month = {apr},
  publisher = {IOP Publishing},
  volume = {36},
  number = {28},
  pages = {283002},
  author = {Yang, Yan-Bin and Wang, Jiong-Hao and Li, Kai and Xu, Yong},
  title = {Higher-order topological phases in crystalline and non-crystalline systems: a review},
  journal = {Journal of Physics: Condensed Matter}
}

@article{pahomi_prr2020,
  title = {{Braiding Majorana corner modes in a second-order topological superconductor}},
  author = {Pahomi, Tudor E. and Sigrist, Manfred and Soluyanov, Alexey A.},
  journal = {Phys. Rev. Res.},
  volume = {2},
  issue = {3},
  pages = {032068},
  numpages = {6},
  year = {2020},
  month = {Sep},
  publisher = {American Physical Society},
  doi = {10.1103/PhysRevResearch.2.032068},
}

@article{wang_prb2024,
  title = {Superconductivity in two-dimensional systems with unconventional {R}ashba bands},
  author = {Wang, Ran and Li, Jiayang and Huang, Xinliang and Wang, Lichuan and Song, Rui and Hao, Ning},
  journal = {Phys. Rev. B},
  volume = {110},
  issue = {13},
  pages = {134517},
  numpages = {14},
  year = {2024},
  month = {Oct},
  publisher = {American Physical Society},
  doi = {10.1103/PhysRevB.110.134517},
}

@article{uday_nature2024,
  author={Uday, Anjana and Lippertz, Gertjan and Moors, Kristof and Legg, Henry F. and Joris, Rikkie and Bliesener, Andrea and Pereira, Lino M. C. and Taskin, A. A. and Ando, Yoichi},
  title={Induced superconducting correlations in a quantum anomalous {H}all insulator},
  journal={Nature Physics},
  year={2024},
  month={Oct},
  day={01},
  volume={20},
  number={10},
  pages={1589-1595},
  issn={1745-2481},
  doi={10.1038/s41567-024-02574-1},
}

@article{wang_prx2012,
  title = {{Simplified Topological Invariants for Interacting Insulators}},
  author = {Wang, Zhong and Zhang, Shou-Cheng},
  journal = {Phys. Rev. X},
  volume = {2},
  issue = {3},
  pages = {031008},
  numpages = {6},
  year = {2012},
  month = {Aug},
  publisher = {American Physical Society},
  doi = {10.1103/PhysRevX.2.031008},
}

@article{kezilebieke_nature2020,
  author={Kezilebieke, Shawulienu and Huda, Md Nurul and Va{\v{n}}o, Viliam and Aapro, Markus and Ganguli, Somesh C. and Silveira, Orlando J. and G{\l}odzik, Szczepan and Foster, Adam S. and Ojanen, Teemu and Liljeroth, Peter},
  title={{Topological superconductivity in a van der Waals heterostructure}},
  journal={Nature},
  year={2020},
  month={Dec},
  day={01},
  volume={588},
  number={7838},
  pages={424-428},
  issn={1476-4687},
  doi={10.1038/s41586-020-2989-y},
}

@article{mazin_prx_2020,
  title = {Ising Superconductivity and Magnetism in ${\mathrm{NbSe}}_{2}$},
  author = {Wickramaratne, Darshana and Khmelevskyi, Sergii and Agterberg, Daniel F. and Mazin, I. I.},
  journal = {Phys. Rev. X},
  volume = {10},
  issue = {4},
  pages = {041003},
  numpages = {13},
  year = {2020},
  month = {Oct},
  publisher = {American Physical Society},
  doi = {10.1103/PhysRevX.10.041003},
}

@article{Roy_2025,
doi = {10.1088/2053-1583/ad7b53},
year = {2024},
month = {oct},
publisher = {IOP Publishing},
volume = {12},
number = {1},
pages = {015004},
author = {Roy, Subhojit and Kreisel, Andreas and Andersen, Brian M and Mukherjee, Shantanu},
title = {Unconventional pairing in {Ising} superconductors: application to monolayer {NbSe$_2$}},
journal = {2D Materials},
}

@article{noat_prb_2015,
  title = {{Quasiparticle spectra of $2\mathrm{H}\ensuremath{-}{\mathrm{NbSe}}_{2}$: Two-band superconductivity and the role of tunneling selectivity}},
  author = {Noat, Y. and Silva-Guill\'en, J. A. and Cren, T. and Cherkez, V. and Brun, C. and Pons, S. and Debontridder, F. and Roditchev, D. and Sacks, W. and Cario, L. and Ordej\'on, P. and Garc\'{\i}a, A. and Canadell, E.},
  journal = {Phys. Rev. B},
  volume = {92},
  issue = {13},
  pages = {134510},
  numpages = {18},
  year = {2015},
  month = {Oct},
  publisher = {American Physical Society},
  doi = {10.1103/PhysRevB.92.134510},
}

@Article{das_npj2023,
  author={Das, S. and Paudyal, H. and Margine, E. R. and Agterberg, D. F. and Mazin, I. I.},
  title={{Electron-phonon coupling and spin fluctuations in the Ising superconductor NbSe$_2$}},
  journal={npj Computational Materials},
  year={2023},
  month={Apr},
  day={26},
  volume={9},
  number={1},
  pages={66},
  issn={2057-3960},
  doi={10.1038/s41524-023-01017-4},
}

@article{giannozzi_qe2009,
  doi = {10.1088/0953-8984/21/39/395502},
  year = {2009},
  month = {sep},
  publisher = {},
  volume = {21},
  number = {39},
  pages = {395502},
  author = {Giannozzi, Paolo and Baroni, Stefano and Bonini, Nicola and Calandra, Matteo and Car, Roberto and Cavazzoni, Carlo and Ceresoli, Davide and Chiarotti, Guido L and Cococcioni, Matteo and Dabo, Ismaila and Dal Corso, Andrea and de Gironcoli, Stefano and Fabris, Stefano and Fratesi, Guido and Gebauer, Ralph and Gerstmann, Uwe and Gougoussis, Christos and Kokalj, Anton and Lazzeri, Michele and Martin-Samos, Layla and Marzari, Nicola and Mauri, Francesco and Mazzarello, Riccardo and Paolini, Stefano and Pasquarello, Alfredo and Paulatto, Lorenzo and Sbraccia, Carlo and Scandolo, Sandro and Sclauzero, Gabriele and Seitsonen, Ari P and Smogunov, Alexander and Umari, Paolo and Wentzcovitch, Renata M},
  title = {{QUANTUM ESPRESSO: a modular and open-source software project for quantum simulations of materials}},
  journal = {Journal of Physics: Condensed Matter}
}

@article{giannozzi_qe2017,
  doi = {10.1088/1361-648X/aa8f79},
  year = {2017},
  month = {oct},
  publisher = {IOP Publishing},
  volume = {29},
  number = {46},
  pages = {465901},
  author = {Giannozzi, P and Andreussi, O and Brumme, T and Bunau, O and Buongiorno Nardelli, M and Calandra, M and Car, R and Cavazzoni, C and Ceresoli, D and Cococcioni, M and Colonna, N and Carnimeo, I and Dal Corso, A and de Gironcoli, S and Delugas, P and DiStasio, R A and Ferretti, A and Floris, A and Fratesi, G and Fugallo, G and Gebauer, R and Gerstmann, U and Giustino, F and Gorni, T and Jia, J and Kawamura, M and Ko, H-Y and Kokalj, A and Küçükbenli, E and Lazzeri, M and Marsili, M and Marzari, N and Mauri, F and Nguyen, N L and Nguyen, H-V and Otero-de-la-Roza, A and Paulatto, L and Poncé, S and Rocca, D and Sabatini, R and Santra, B and Schlipf, M and Seitsonen, A P and Smogunov, A and Timrov, I and Thonhauser, T and Umari, P and Vast, N and Wu, X and Baroni, S},
  title = {{Advanced capabilities for materials modelling with Quantum ESPRESSO}},
  journal = {Journal of Physics: Condensed Matter}
}

@article{giann_jcp_2020,
    author = {Giannozzi, Paolo and Baseggio, Oscar and Bonfà, Pietro and Brunato, Davide and Car, Roberto and Carnimeo, Ivan and Cavazzoni, Carlo and de Gironcoli, Stefano and Delugas, Pietro and Ferrari Ruffino, Fabrizio and Ferretti, Andrea and Marzari, Nicola and Timrov, Iurii and Urru, Andrea and Baroni, Stefano},
    title = {Quantum ESPRESSO toward the exascale},
    journal = {The Journal of Chemical Physics},
    volume = {152},
    number = {15},
    pages = {154105},
    year = {2020},
    month = {04},
    issn = {0021-9606},
    doi = {10.1063/5.0005082},
}

@article{sab_prb_2013,
  title = {Nonlocal van der Waals density functional made simple and efficient},
  author = {Sabatini, Riccardo and Gorni, Tommaso and de Gironcoli, Stefano},
  journal = {Phys. Rev. B},
  volume = {87},
  issue = {4},
  pages = {041108},
  numpages = {4},
  year = {2013},
  month = {Jan},
  publisher = {American Physical Society},
  doi = {10.1103/PhysRevB.87.041108},
}

@article{v_prl_2009,
  title = {Nonlocal van der Waals Density Functional Made Simple},
  author = {Vydrov, Oleg A. and Van Voorhis, Troy},
  journal = {Phys. Rev. Lett.},
  volume = {103},
  issue = {6},
  pages = {063004},
  numpages = {4},
  year = {2009},
  month = {Aug},
  publisher = {American Physical Society},
  doi = {10.1103/PhysRevLett.103.063004},
}

@article{vb_ssp_1972,
doi = {10.1088/0022-3719/5/13/012},
year = {1972},
month = {jul},
publisher = {},
volume = {5},
number = {13},
pages = {1629},
author = {U von Barth and L Hedin},
title = {A local exchange-correlation potential for the spin polarized case. i},
journal = {Journal of Physics C: Solid State Physics},
}

@article{oncvpsp_prb2013,
  title = {{Optimized norm-conserving Vanderbilt pseudopotentials}},
  author = {Hamann, D. R.},
  journal = {Phys. Rev. B},
  volume = {88},
  issue = {8},
  pages = {085117},
  numpages = {10},
  year = {2013},
  month = {Aug},
  publisher = {American Physical Society},
  doi = {10.1103/PhysRevB.88.085117},
}

@book{zhu_bdgbook,
  author = {Zhu, Jian-Xin},
  address = {Cham},
  isbn = {9783319313146},
  issn = {0075-8450},
  publisher = {Springer Nature},
  series = {Lecture Notes in Physics},
  title = {{Bogoliubov-de Gennes Method and Its Applications}},
  volume = {924},
  year = {2016},
}

@article{cuperus_scipost2025,
	title={{Non-topological edge-localized Yu-Shiba-Rusinov states in CrBr$_3$/NbSe$_2$ heterostructures}},
	author={Jan P. Cuperus and Daniel Vanmaekelbergh and Ingmar Swart},
	journal={SciPost Phys.},
	volume={18},
	pages={141},
	year={2025},
	publisher={SciPost},
	doi={10.21468/SciPostPhys.18.4.141},
}

@Article{li_natcomm2024,
author={Li, Yuanji and Yin, Ruotong and Li, Mingzhe and Gong, Jiashuo and Chen, Ziyuan and Zhang, Jiakang and Yan, Ya-Jun and Feng, Dong-Lai},
  title={{Observation of Yu-Shiba-Rusinov-like states at the edge of CrBr$_3$/NbSe$_2$ heterostructure}},
  journal={Nature Communications},
  year={2024},
  month={Nov},
  day={22},
  volume={15},
  number={1},
  pages={10121},
  issn={2041-1723},
  doi={10.1038/s41467-024-54525-2},
}

@article{bansil_prb2023,
  title = {{Atomistic modeling of a superconductor--transition metal dichalcogenide--superconductor Josephson junction}},
  author = {Nieminen, Jouko and Dhara, Sayandip and Chiu, Wei-Chi and Mucciolo, Eduardo R. and Bansil, Arun},
  journal = {Phys. Rev. B},
  volume = {107},
  issue = {17},
  pages = {174524},
  numpages = {16},
  year = {2023},
  month = {May},
  publisher = {American Physical Society},
  doi = {10.1103/PhysRevB.107.174524},
}

@article{kezilebieke_nanolett2022,
  author={Kezilebieke, Shawulienu and Va{\v{n}}o, Viliam and Huda, Md N. and Aapro, Markus and Ganguli, Somesh C. and Liljeroth, Peter and Lado, Jose L.},
  title={Moir{\'e}-{Enabled Topological Superconductivity}},
  journal={Nano Letters},
  year={2022},
  month={Jan},
  day={12},
  publisher={American Chemical Society},
  volume={22},
  number={1},
  pages={328-333},
  issn={1530-6984},
  doi={10.1021/acs.nanolett.1c03856},
}

@article{kezilebieke_adma2021,
  author = {Kezilebieke, Shawulienu and Silveira, Orlando J. and Huda, Md N. and Vaňo, Viliam and Aapro, Markus and Ganguli, Somesh Chandra and Lahtinen, Jouko and Mansell, Rhodri and van Dijken, Sebastiaan and Foster, Adam S. and Liljeroth, Peter},
  title = {{Electronic and Magnetic Characterization of Epitaxial CrBr$_3$ Monolayers on a Superconducting Substrate}},
  journal = {Advanced Materials},
  volume = {33},
  number = {23},
  pages = {2006850},
  doi = {https://doi.org/10.1002/adma.202006850},
  year = {2021}
}

@article{Souza2001,
  title = {{Maximally localized Wannier functions for entangled energy bands}},
  author = {Souza, Ivo and Marzari, Nicola and Vanderbilt, David},
  journal = {Phys. Rev. B},
  volume = {65},
  issue = {3},
  pages = {035109},
  numpages = {13},
  year = {2001},
  month = {Dec},
  publisher = {American Physical Society},
  doi = {10.1103/PhysRevB.65.035109},
}

@article{marzari_prb_1997,
  title = {{Maximally localized generalized Wannier functions for composite energy bands}},
  author = {Marzari, Nicola and Vanderbilt, David},
  journal = {Phys. Rev. B},
  volume = {56},
  issue = {20},
  pages = {12847--12865},
  numpages = {0},
  year = {1997},
  month = {Nov},
  publisher = {American Physical Society},
  doi = {10.1103/PhysRevB.56.12847},
}

@article{Mostofi2008,
  title = {{Wannier90: A tool for obtaining maximally-localised Wannier functions}},
  journal = "Comput. Phys. Commun.",
  volume = "178",
  number = "9",
  pages = "685 - 699",
  year = "2008",
  issn = "0010-4655",
  doi = "https://doi.org/10.1016/j.cpc.2007.11.016",
  author = "Arash A. Mostofi and Jonathan R. Yates and Young-Su Lee and Ivo Souza and David Vanderbilt and Nicola Marzari"
}

@article{Pizzi2020,
	author = {Giovanni Pizzi and Valerio Vitale and Ryotaro Arita and Stefan Bl{\"u}gel and Frank Freimuth and Guillaume G{\'{e}}ranton and Marco Gibertini and Dominik Gresch and Charles Johnson and Takashi Koretsune and Julen Iba{\~{n}}ez-Azpiroz and Hyungjun Lee and Jae-Mo Lihm and Daniel Marchand and Antimo Marrazzo and Yuriy Mokrousov and Jamal I Mustafa and Yoshiro Nohara and Yusuke Nomura and Lorenzo Paulatto and Samuel Ponc{\'{e}} and Thomas Ponweiser and Junfeng Qiao and Florian Th{\"o}le and Stepan S Tsirkin and Ma{\l}gorzata Wierzbowska and Nicola Marzari and David Vanderbilt and Ivo Souza and Arash A Mostofi and Jonathan R Yates},
	doi = {10.1088/1361-648x/ab51ff},
	journal = {J. Phys. Condens. Matter},
	month = {jan},
	number = {16},
	pages = {165902},
	publisher = {{IOP} Publishing},
	title = {{Wannier90 as a community code: new features and applications}},
	volume = {32},
	year = 2020,
}

@book{Vanderbilt2018,
  author = {Vanderbilt, D.},
  isbn = {978-1-107-15765-1},
  lccn = {2018018455},
  publisher = {Cambridge University Press},
  title = {{Berry Phases in Electronic Structure Theory: Electric Polarization, Orbital Magnetization and Topological Insulators}},
  url = {https://books.google.ch/books?id=485FtgEACAAJ},
  year = {2018},
}

@article{soluyanov_prb2011,
  title = {Computing topological invariants without inversion symmetry},
  author = {Soluyanov, Alexey A. and Vanderbilt, David},
  journal = {Phys. Rev. B},
  volume = {83},
  issue = {23},
  pages = {235401},
  numpages = {9},
  year = {2011},
  month = {Jun},
  publisher = {American Physical Society},
  doi = {10.1103/PhysRevB.83.235401},
}

@article{gresch_prb2017,
  title = {{Z2Pack: Numerical implementation of hybrid Wannier centers for identifying topological materials}},
  author = {Gresch, Dominik and Aut\`es, Gabriel and Yazyev, Oleg V. and Troyer, Matthias and Vanderbilt, David and Bernevig, B. Andrei and Soluyanov, Alexey A.},
  journal = {Phys. Rev. B},
  volume = {95},
  issue = {7},
  pages = {075146},
  numpages = {24},
  year = {2017},
  month = {Feb},
  publisher = {American Physical Society},
  doi = {10.1103/PhysRevB.95.075146},
}

@article{bergeret_rmp2005,
  title = {Odd triplet superconductivity and related phenomena in superconductor-ferromagnet structures},
  author = {Bergeret, F. S. and Volkov, A. F. and Efetov, K. B.},
  journal = {Rev. Mod. Phys.},
  volume = {77},
  issue = {4},
  pages = {1321--1373},
  numpages = {0},
  year = {2005},
  month = {Nov},
  publisher = {American Physical Society},
  doi = {10.1103/RevModPhys.77.1321},
}

@Inbook{efetov_springer2008,
  author={Efetov, Konstantin B. and Garifullin, Ilgiz A. and Volkov, Anatoly F. and Westerholt, Kurt},
  editor={Zabel, Hartmut and Bader, Samuel D.},
  title={{Proximity Effects in Ferromagnet/Superconductor Heterostructures}},
  bookTitle={Magnetic Heterostructures: Advances and Perspectives in Spinstructures and Spintransport},
  year={2008},
  publisher={Springer Berlin Heidelberg},
  address={Berlin, Heidelberg},
  pages={251--290},
  isbn={978-3-540-73462-8},
  doi={10.1007/978-3-540-73462-8_5},
}

@article{buzdin_rmp2005,
  title = {Proximity effects in superconductor-ferromagnet heterostructures},
  author = {Buzdin, A. I.},
  journal = {Rev. Mod. Phys.},
  volume = {77},
  issue = {3},
  pages = {935--976},
  numpages = {0},
  year = {2005},
  month = {Sep},
  publisher = {American Physical Society},
  doi = {10.1103/RevModPhys.77.935},
}

@article{damle2015,
  author={Damle, Anil and Lin, Lin and Ying, Lexing},
  title={{Compressed Representation of Kohn--Sham Orbitals via Selected Columns of the Density Matrix}},
  journal={Journal of Chemical Theory and Computation},
  year={2015},
  month={Apr},
  day={14},
  publisher={American Chemical Society},
  volume={11},
  number={4},
  pages={1463-1469},
  issn={1549-9618},
  doi={10.1021/ct500985f},
}

@article{damle2017,
title = {{SCDM-k: Localized orbitals for solids via selected columns of the density matrix}},
journal = {Journal of Computational Physics},
volume = {334},
pages = {1-15},
year = {2017},
issn = {0021-9991},
doi = {https://doi.org/10.1016/j.jcp.2016.12.053},
author = {Anil Damle and Lin Lin and Lexing Ying}
}

@article{vansetten_cpc2018,
  title = {{The PseudoDojo: Training and grading a 85 element optimized norm-conserving pseudopotential table}},
  journal = {Computer Physics Communications},
  volume = {226},
  pages = {39-54},
  year = {2018},
  issn = {0010-4655},
  doi = {https://doi.org/10.1016/j.cpc.2018.01.012},
  author = {M.J. {van Setten} and M. Giantomassi and E. Bousquet and M.J. Verstraete and D.R. Hamann and X. Gonze and G.-M. Rignanese}
}

@article{sssplib,
   title={Precision and efficiency in solid-state pseudopotential calculations},
   author={Prandini, Gianluca and Marrazzo, Antimo and Castelli, Ivano E and Mounet, Nicolas and Marzari, Nicola},
   journal={npj Computational Materials},
   volume={4},
   number={1},
   pages={72},
   year={2018},
   issn = {2057-3960},
   doi = {10.1038/s41524-018-0127-2},
   note = {\href{http://materialscloud.org/sssp}{http://materialscloud.org/sssp}},
   publisher={Nature Publishing Group UK London}
}

@article{garrity_cms2014,
  title = {{Pseudopotentials for high-throughput DFT calculations}},
  journal = {Computational Materials Science},
  volume = {81},
  pages = {446-452},
  year = {2014},
  issn = {0927-0256},
  doi = {https://doi.org/10.1016/j.commatsci.2013.08.053},
  author = {Kevin F. Garrity and Joseph W. Bennett and Karin M. Rabe and David Vanderbilt}
}

@article{calzolari_prb2004,
  title = {Ab initio transport properties of nanostructures from maximally localized Wannier functions},
  author = {Calzolari, Arrigo and Marzari, Nicola and Souza, Ivo and Buongiorno Nardelli, Marco},
  journal = {Phys. Rev. B},
  volume = {69},
  issue = {3},
  pages = {035108},
  numpages = {10},
  year = {2004},
  month = {Jan},
  publisher = {American Physical Society},
  doi = {10.1103/PhysRevB.69.035108},
}

@article{ToschiPRB2005,
  title = {{Energetic balance of the superconducting transition across the BCS---Bose Einstein crossover in the attractive Hubbard model}},
  author = {Toschi, A. and Capone, M. and Castellani, C.},
  journal = {Phys. Rev. B},
  volume = {72},
  issue = {23},
  pages = {235118},
  numpages = {10},
  year = {2005},
  month = {Dec},
  publisher = {American Physical Society},
  doi = {10.1103/PhysRevB.72.235118},
}

@article{GeorgesRMP1996,
  title = {Dynamical mean-field theory of strongly correlated fermion systems and the limit of infinite dimensions},
  author = {Georges, Antoine and Kotliar, Gabriel and Krauth, Werner and Rozenberg, Marcelo J.},
  journal = {Rev. Mod. Phys.},
  volume = {68},
  issue = {1},
  pages = {13--125},
  numpages = {0},
  year = {1996},
  month = {Jan},
  publisher = {American Physical Society},
  doi = {10.1103/RevModPhys.68.13},
}

@article{Nomura2015,
  author = {Yusuke Nomura  and Shiro Sakai  and Massimo Capone  and Ryotaro Arita },
  title = {{Unified understanding of superconductivity and Mott transition in alkali-doped fullerides from first principles}},
  journal = {Science Advances},
  volume = {1},
  number = {7},
  pages = {e1500568},
  year = {2015},
  doi = {10.1126/sciadv.1500568},
}

@article{Capone2009,
  title = {Colloquium: {M}odeling the unconventional superconducting properties of expanded ${A}_{3}\mathrm{C}_{60}$ fullerides},
  author = {Capone, Massimo and Fabrizio, Michele and Castellani, Claudio and Tosatti, Erio},
  journal = {Rev. Mod. Phys.},
  volume = {81},
  issue = {2},
  pages = {943--958},
  numpages = {0},
  year = {2009},
  month = {Jun},
  publisher = {American Physical Society},
  doi = {10.1103/RevModPhys.81.943},
}

@article{BacqLabrueil2025,
  title = {{Toward an Ab Initio Theory of High-Temperature Superconductors: A Study of Multilayer Cuprates}},
  author = {Bacq-Labreuil, Benjamin and Lacasse, Benjamin and Tremblay, A.-M. S. and S\'en\'echal, David and Haule, Kristjan},
  journal = {Phys. Rev. X},
  volume = {15},
  issue = {2},
  pages = {021071},
  numpages = {44},
  year = {2025},
  month = {May},
  publisher = {American Physical Society},
  doi = {10.1103/PhysRevX.15.021071},
}

@Article{Kitatani2020,
  author={Kitatani, Motoharu and Si, Liang and Janson, Oleg and Arita, Ryotaro and Zhong, Zhicheng and Held, Karsten},
  title={Nickelate superconductors---a renaissance of the one-band {H}ubbard model},
  journal={npj Quantum Materials},
  year={2020},
  month={Aug},
  day={21},
  volume={5},
  number={1},
  pages={59},
  issn={2397-4648},
  doi={10.1038/s41535-020-00260-y},
}

@article{Berg2008,
  title = {Route to high-temperature superconductivity in composite systems},
  author = {Berg, Erez and Orgad, Dror and Kivelson, Steven A.},
  journal = {Phys. Rev. B},
  volume = {78},
  issue = {9},
  pages = {094509},
  numpages = {7},
  year = {2008},
  month = {Sep},
  publisher = {American Physical Society},
  doi = {10.1103/PhysRevB.78.094509},
}

@article{Mazza2021,
  title = {Interface and bulk superconductivity in superconducting heterostructures with enhanced critical temperatures},
  author = {Mazza, Giacomo and Amaricci, Adriano and Capone, Massimo},
  journal = {Phys. Rev. B},
  volume = {103},
  issue = {9},
  pages = {094514},
  numpages = {8},
  year = {2021},
  month = {Mar},
  publisher = {American Physical Society},
  doi = {10.1103/PhysRevB.103.094514},
}

@article{Smit2025,
  author = {Smit, Stef et al},
  eprint = {2506.01448},
  month = {jan},
  Journal = {ArXiv:2506.01448 },
  title = {{Enhanced coherence and layer-selective charge order in a trilayer cuprate superconductor}},
  year = {2025}
}

@article{Sous2023,
  title = {{Bipolaronic High-Temperature Superconductivity}},
  author = {Zhang, C. and Sous, J. and Reichman, D. R. and Berciu, M. and Millis, A. J. and Prokof'ev, N. V. and Svistunov, B. V.},
  journal = {Phys. Rev. X},
  volume = {13},
  issue = {1},
  pages = {011010},
  numpages = {19},
  year = {2023},
  month = {Jan},
  publisher = {American Physical Society},
  doi = {10.1103/PhysRevX.13.011010},
}

@article{Capone2003,
  title = {{Polaron Crossover and Bipolaronic Metal-Insulator Transition in the Half-Filled Holstein Model}},
  author = {Capone, M. and Ciuchi, S.},
  journal = {Phys. Rev. Lett.},
  volume = {91},
  issue = {18},
  pages = {186405},
  numpages = {4},
  year = {2003},
  month = {Oct},
  publisher = {American Physical Society},
  doi = {10.1103/PhysRevLett.91.186405},
}

@article{Micnas1990,
  title = {Superconductivity in narrow-band systems with local nonretarded attractive interactions},
  author = {Micnas, R. and Ranninger, J. and Robaszkiewicz, S.},
  journal = {Rev. Mod. Phys.},
  volume = {62},
  issue = {1},
  pages = {113--171},
  numpages = {0},
  year = {1990},
  month = {Jan},
  publisher = {American Physical Society},
  doi = {10.1103/RevModPhys.62.113},
}

@article{Zhang2023,
  title = {{Bipolaronic High-Temperature Superconductivity}},
  author = {Zhang, C. and Sous, J. and Reichman, D. R. and Berciu, M. and Millis, A. J. and Prokof'ev, N. V. and Svistunov, B. V.},
  journal = {Phys. Rev. X},
  volume = {13},
  issue = {1},
  pages = {011010},
  numpages = {19},
  year = {2023},
  month = {Jan},
  publisher = {American Physical Society},
  doi = {10.1103/PhysRevX.13.011010},
}

@article{Chakraverty1985,
  author = {B. K. Chakraverty and J. Ranninger},
  title = {Bipolarons and superconductivity},
  journal = {Philosophical Magazine B},
  volume = {52},
  number = {3},
  pages = {669--678},
  year = {1985},
  publisher = {Taylor \& Francis},
  doi = {10.1080/13642818508240628},
}

@article{Witt2024,
  title={{Bypassing the lattice BCS--BEC crossover in strongly correlated superconductors through multiorbital physics}},
  author={Witt, Niklas and Nomura, Yusuke and Brener, Sergey and Arita, Ryotaro and Lichtenstein, Alexander I and Wehling, Tim O},
  journal={npj Quantum Materials},
  volume={9},
  number={1},
  pages={1--10},
  year={2024},
  publisher={Nature Publishing Group},
  doi={10.1038/s41535-024-00706-7},
}

@article{li_prl2026,
  title = {{Spin-Polarized Josephson Supercurrent in Nodeless Altermagnets}},
  author = {Li, Chuang and Hou, Jin-Xing and Zhang, Fu-Chun and Zhang, Song-Bo and Hu, Lun-Hui},
  journal = {Phys. Rev. Lett.},
  volume = {136},
  issue = {11},
  pages = {116701},
  numpages = {8},
  year = {2026},
  month = {Mar},
  publisher = {American Physical Society},
  doi = {10.1103/ds5z-fxy5},
}

@misc{chen_arxiv2025,
      title={{Intertwining Josephson and Vortex Topologies in Conventional Superconductors}}, 
      author={Zhuo Chen and Jiangxu Li and Lun-Hui Hu and Zhen Bi and Rui-Xing Zhang},
      year={2025},
      eprint={2504.07066},
      archivePrefix={arXiv},
      primaryClass={cond-mat.supr-con},
}

@article{ouassou_prl2023,
  title = {{dc Josephson Effect in Altermagnets}},
  author = {Ouassou, Jabir Ali and Brataas, Arne and Linder, Jacob},
  journal = {Phys. Rev. Lett.},
  volume = {131},
  issue = {7},
  pages = {076003},
  numpages = {6},
  year = {2023},
  month = {Aug},
  publisher = {American Physical Society},
  doi = {10.1103/PhysRevLett.131.076003},
}

@book{bernevig_book2013,
	title = {Topological {Insulators} and {Topological} {Superconductors}},
	isbn = {978-1-4008-4673-3},
	doi = {10.1515/9781400846733},
	urldate = {2023-08-31},
	publisher = {Princeton University Press},
	author = {Bernevig, B. Andrei and Hughes, Taylor L.},
	month = {Apr},
	year = {2013},
}

@book{stanescu_book2024,
  author = {Stanescu, Tudor D.},
  title = {{Introduction to Topological Quantum Matter \& Quantum Computation}},
  publisher = {CRC Press},
  year = {2024},
  doi = {10.1201/9781003226048}
}

@article{belzig_prb1996,
  title = {Local density of states in a dirty normal metal connected to a superconductor},
  author = {Belzig, W. and Bruder, C. and Sch\"on, Gerd},
  journal = {Phys. Rev. B},
  volume = {54},
  issue = {13},
  pages = {9443--9448},
  numpages = {0},
  year = {1996},
  month = {Oct},
  publisher = {American Physical Society},
  doi = {10.1103/PhysRevB.54.9443},
}

@article{hui_prb2015,
  title = {Bulk disorder in the superconductor affects proximity-induced topological superconductivity},
  author = {Hui, Hoi-Yin and Sau, Jay D. and Das Sarma, S.},
  journal = {Phys. Rev. B},
  volume = {92},
  issue = {17},
  pages = {174512},
  numpages = {11},
  year = {2015},
  month = {Nov},
  publisher = {American Physical Society},
  doi = {10.1103/PhysRevB.92.174512},
}

@article{materialscloud_data,
  author = {{Baù Nicolas} and {Dowlatabadi Mitra} and {Chiarotti Tommaso} and {Capone Massimo} and {Marrazzo  Antimo}},
  title = {First-principles real-space embedding theory of the superconducting proximity effect},
  volume ={2026.133},
  journal = {Materials Cloud Archive},
  year = {2026},
  doi = {10.24435/materialscloud:wv-cm},
}
\end{document}